\documentclass[12pt,a4paper]{article}
\usepackage[left=2cm,right=2cm,top=2cm,bottom=2cm]{geometry}
\usepackage{amsmath,amssymb}
\numberwithin{equation}{section}
\usepackage{hyperref}
\usepackage{graphicx}
\usepackage{booktabs}
\usepackage{color,soul}
\usepackage[utf8]{inputenc}
\usepackage{multirow}
\usepackage{authblk}
\usepackage{lscape}

\usepackage[font=small,labelfont=bf]{caption}

\usepackage[
    backend=biber,
    sortcites=true,
    style=nature,
    maxbibnames=99,
]{biblatex}

\AtEveryBibitem{%
 \clearfield{url}%
 \clearfield{urlyear}%
 \clearfield{langid}%
 \clearfield{month}%
 \clearfield{day}%
 \clearfield{issn}%
 \clearfield{isbn}%
 \clearfield{doi}
 \clearfield{pagetotal}
 \clearfield{ISSN}
 \clearfield{tex.ids}
 \clearfield{note}
 \clearfield{eprinttype}%
 \clearfield{eprint}
 \clearlist{language}%
 \clearfield{language}%
 \clearlist{editor}%
 \clearfield{editor}%
}

\usepackage{xr} 
\makeatletter
\newcommand*{\addFileDependency}[1]{
\typeout{(#1)}
\@addtofilelist{#1}
\IfFileExists{#1}{}{\typeout{No file #1.}}
}\makeatother

\newcommand*{\myexternaldocument}[1]{%
\externaldocument{#1}%
\addFileDependency{#1.tex}%
\addFileDependency{#1.aux}%
}

\myexternaldocument{supp}

\usepackage[colorinlistoftodos, textwidth=36mm, textsize=singlespacetiny, shadow]{todonotes}

\usepackage{setspace}
\def\methods{\textit{Materials \& Methods}}
\def\appendix{\textit{Supporting Information}}

\newcommand{\beginsupplement}{%
        \setcounter{table}{0}
        \renewcommand{\thetable}{S\arabic{table}}%
        \setcounter{figure}{0}
        \renewcommand{\thefigure}{S\arabic{figure}}%
     }

\begin{document}

\title{The origin, consequence, and visibility of criticism in science}

\author[1,a]{Bingsheng Chen}
\author[1,a]{Dakota Murray}
\author[a]{Yixuan Liu}
\author[a,b,c]{Albert-L\'{a}szl\'{o}  Barab\'{a}si }
\affil[a]{Network Science Institute, Northeastern University, Boston, MA, 02115}
\affil[b]{Department of Medicine, Brigham and Women’s Hospital, Harvard Medical School, Boston, MA, 02115}
\affil[c]{Department of Network and Data Science, Central European University, Budapest 1051, Hungary}
\affil[1]{Authors contributed equally to this work}


\maketitle

\clearpage
\begin{abstract}
Critique between peers plays a vital role in the production of scientific knowledge.
Yet, there is limited empirical evidence on the origins of criticism, its effects on the papers and individuals involved, and its visibility within the scientific literature. Here, we address these gaps through a data-driven analysis of papers that received substantiated and explicit written criticisms.
Our analysis draws on data representing over 3,000 ``critical letters''---papers explicitly published to critique another---from four high profile journals, with each letter linked to its target paper.
We find that the papers receiving critical letters are disproportionately among the most highly-cited in their respective journal and, to a lesser extent, among the most interdisciplinary and novel.
However, despite the theoretical importance of criticism in scientific progress, we observe no evidence that receiving a critical letter affects a paper's citation trajectory or the productivity and citation impact of its authors.
One explanation for the limited consequence of critical letters is that they often go unnoticed. 
Indeed, we find that critical letters attract only a small fraction of the citations received by their targets, even years after publication.
An analysis of topical similarity between criticized papers and their citing papers indicates that critical letters are primarily cited by researchers actively engaged in a similar field of study, whereas they are overlooked by more distant communities.
Although criticism is celebrated as a cornerstone to science, our findings reveal that it is concentrated on high-impact papers, has minimal measurable consequences, and suffers from limited visibility.
These results raise important questions about the role and value of critique in scientific practice.
\end{abstract}

\section*{Introduction}
Constructive criticism between peers is viewed as essential to the functioning and flourishing of the scientific process.
It drives scientists to pursue new avenues of research, reconcile conflicting worldviews, remedy errors and biases, and can even lay the foundation for revolutionary new theories~\cite{decruz_2013_value,bastian2014stronger}. 
It defines key moments in the history of science such as the Galileo affair~\cite{feyerabend_1975_against}, Einstein’s and Bohr’s divergence over quantum theory~\cite{kumar_2010_quantum}, and the taxonomic classification of \textit{homo floresiensis}~\cite{decruz_2013_value}.
Epistemological theories of science consistently emphasize the central role of criticism, often using terms such as dissent, controversy, disagreement, or debate~\cite{feyerabend_1975_against, kitcher_advancement_1993,latour_science_1988}; famously, Karl Popper's theory of falsification positions criticism of scientific ideas as the root driver of progress and the key characteristic that differentiates science from pseudoscience~\cite{popper_logic_1959}.
In light of the perceived benefits of criticism, some scholars have advocated for the broader adoption and formalization of critical practices, including post-publication peer review~\cite{bastian2014stronger,winker2015promise}.

Despite its reputation as both vital and ubiquitous to science, there is little empirical evidence documenting the origins and effects of scientific criticism.
Much of the existing research has focused on criticism from non-scientists or biased actors seeking to undermine scientific consensus for political or economic purposes~\cite{oreskes_2010_merchants} and the public perception of debates between scientists~\cite{deroover_2022_experts}.
This body of work often examines public controversies arising from such efforts, even when the underlying scientific questions have long been settled within the scientific community~\cite{lewandowsky_2012_consensus}.
In contrast, this study investigates the less-explored domain of disagreements between epistemic peers---scientists with comparable expertise and access to evidence---who engage in good-faith, substantiated critique~\cite{christensen_2007_disagreement}.
Our focus excludes retractions, where a study’s credibility is discredited due to malpractice, fraud, or major errors. Instead, we examine narrower critiques, often targeting a study’s methodology or interpretation. 
These critiques stop short of outright condemnation but may introduce uncertainty about a study’s credibility or significance that lead to the community re-evaluating the targeted paper.

Existing studies on criticism in science have primarily focused on cataloging its incidence and the presence of debate in published texts, often through analyses of in-text citations~\cite{lamers_2021_disagreement,catalini_2015_negative} or through literature surveys~\cite{cook_2013_quantifying}.
For papers, such work has observed that criticism is disproportionately leveled against those with the highest impact~\cite{catalini_2015_negative,radicchi_2012_comment}.
Outside of citation impact, however, not much is known about the distinguishing characteristics of papers targeted by criticism.
For careers, the consequences of criticism are only understood for its most severe consequences, retractions, which represent only a fraction of all criticism~\cite{azoulay_2017_scandal,peng2022retracted}.

We investigate three key questions about criticism in science: (\textbf{1}) what distinguishes papers that receive criticism; (\textbf{2}) what are the consequences of receiving criticism for the paper and the authors involved; and (\textbf{3}) to what extent are critical letters seen and cited? 
Criticism is operationalized using a type of document we term \textit{critical letters}---a form of post-publication peer review hosted by many journals~\cite{hardwicke_2022_critique} under names such as ``comment'', ``letter to the editor'' or ``matters arising''.
These documents are published with the explicit goal of critiquing recent work.
Critical letters typically undergo editorial or peer review and are often accompanied by a response from the authors of the criticized publication.
While not representative of all forms of criticism, our focus on critical letters offers several advantages:
(i) having undergone editorial review, critical letters are likely to represent substantiated and non-trivial critiques between peers;
(ii) critical letters are, by and large, explicit and unambiguous instances of criticism, in contrast to the often subtle and nuanced critiques embedded in other forms of academic writing;
(iii) because we focus on elite journals, these critical letters are high-profile, making them the best-case scenario for visibility.
As such, critical letters provide an ideal case study: if scholarly criticism has measurable consequences, we should expect to observe them in this context.

Critical letters were sourced from several prominent, widely-read, and disciplinary-diverse journals in which they are routinely published: \textit{Nature}, \textit{Science}, \textit{Proceedings of the National Academy of Sciences} (PNAS), \textit{Physical Review Letters }(PRL), and \textit{Physical Review A} through \textit{Physical Review E} (collapsed into a single category labeled ``\textit{Other APS}'', unless otherwise noted).
Other journals were examined but ultimately excluded due to their letters lacking substance, not explicitly presenting criticism, not being prominent enough, or having too few letters for robust analysis (see \appendix).
We identified 595 critical letters and 517 targeted publications between 2007 and 2019 from \textit{PNAS}, 238 letters and 185 targeted publications between 2004 and 2019 from \textit{Nature}, 480 letters and 404 targeted publications between 2003 and 2019 from \textit{Science}, and 1,831 letters and 1,682 targeted publications in \textit{PRL} journals between 1990 and 2019 (see \methods).
We manually annotated a sample of recent letters published in \textit{Nature}, \textit{Science}, and \textit{PNAS}, finding the majority to be valid instances of criticism in each of the three journals, with the lowest (65\%) in \textit{PNAS}.

%
%
\section*{Results}

\subsection*{The origins of criticism}

Why are some papers the subject of formal criticism while others are not?
To address this question, we investigated how papers that received criticism differ from others published in the same journal that did not.
Specifically, we focused on three theoretically significant characteristics: the paper’s attention, its interdisciplinarity, and its novelty.

The more attention that a paper receives, the more likely it is to attract critical scrutiny, increasing the chances that readers may identify issues and submit a critical letter. Prior research shows that criticism tends to cluster among the most highly cited papers~\cite{radicchi_2012_comment, catalini_2015_negative},  and this attention-driven scrutiny may help explain the disproportionate rate of retractions in high-impact journals~\cite{cokol_2007_retracted}.
Moreover, journal editors, aiming to optimize readership within limited issue space, may prioritize publishing critiques of the most influential papers. Here, we adopt a bibliometric conceptualization of attention, treating it as equivalent to impact, and operationalizing it through the proxy of citations accumulated within 3 years of a paper's publication.

\begin{figure}
    \centering
    \includegraphics[width=0.85\linewidth]{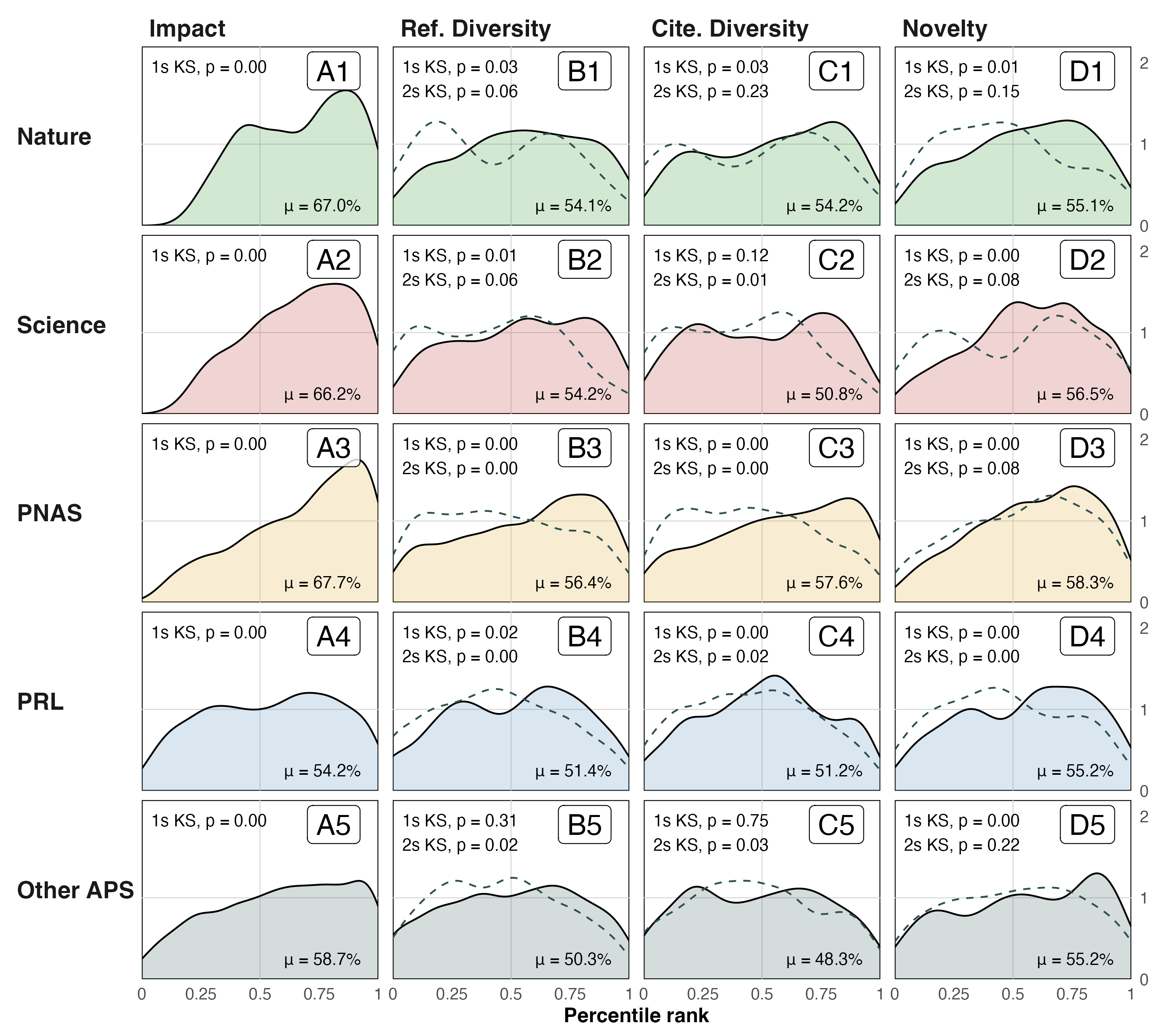}
    \caption{
    \textbf{Characterizing the population of papers targeted by critical letters}.
    Shown is the distribution of percentile ranks for publications targeted by critical letters across four metrics and journals
    Percentile ranks are calculated within each journal, over a four-year time period, and within a high-level field. 
    The metrics include \textbf{(A.1-5)} citations received by papers within 3 years of publication;
    \textbf{(B.1-5)}
    the diversity of referenced papers, measured using the Simpson's Index of their high-level field categories;
    \textbf{(C.1-5)}
    the diversity of citing papers published within five years of the target, calculated using the same index;
    \textbf{(D.1-5)}
    a bibliometric indicator measuring the atypicality of a paper’s cited references.
    For detailed definitions, see \methods. 
    For each journal and metric, $\mu$ (referred to in-text as $mu_{rank}$ denotes the average percentile rank of papers targeted by criticism.
    To examine whether differences in diversity and novelty are mediated by citation impact, a second population of matched papers (black dashed line) is included. Using propensity score matching, we identify for each targeted paper the nearest match from the same journal, in the same high-level field, published within a three-year window, and within a 5\% tolerance of logarithmic three-year impact (see \methods for details).
    To guide interpretation, we employ two Kolmogorov–Smirnov (KS) tests.
    (\textit{``1s KS''}) A one-sample, one-sided KS test compares the distribution of percentile ranks for targeted papers against a uniform distribution. A low p-value indicates that targeted papers are concentrated among higher ranks, rather than being randomly distributed across the journal.
    (\textit{``2s KS''}) A two-sample, one-sided KS test compares the distribution of percentile ranks for targeted papers against the matched population. A low p-value suggests that targeted papers are concentrated among higher ranks compared to their matched counterparts.
    P-values for each test are shown in each panel, and results are shown in greater detail in Tables~\ref*{si:table:paper-features-1sKS}-\ref*{si:table:paper-features-2sKS}.
    }
    \label{fig:paper-features}
\end{figure}

\begin{figure}[p!]
    \centering
    \includegraphics[width=1.0\linewidth]{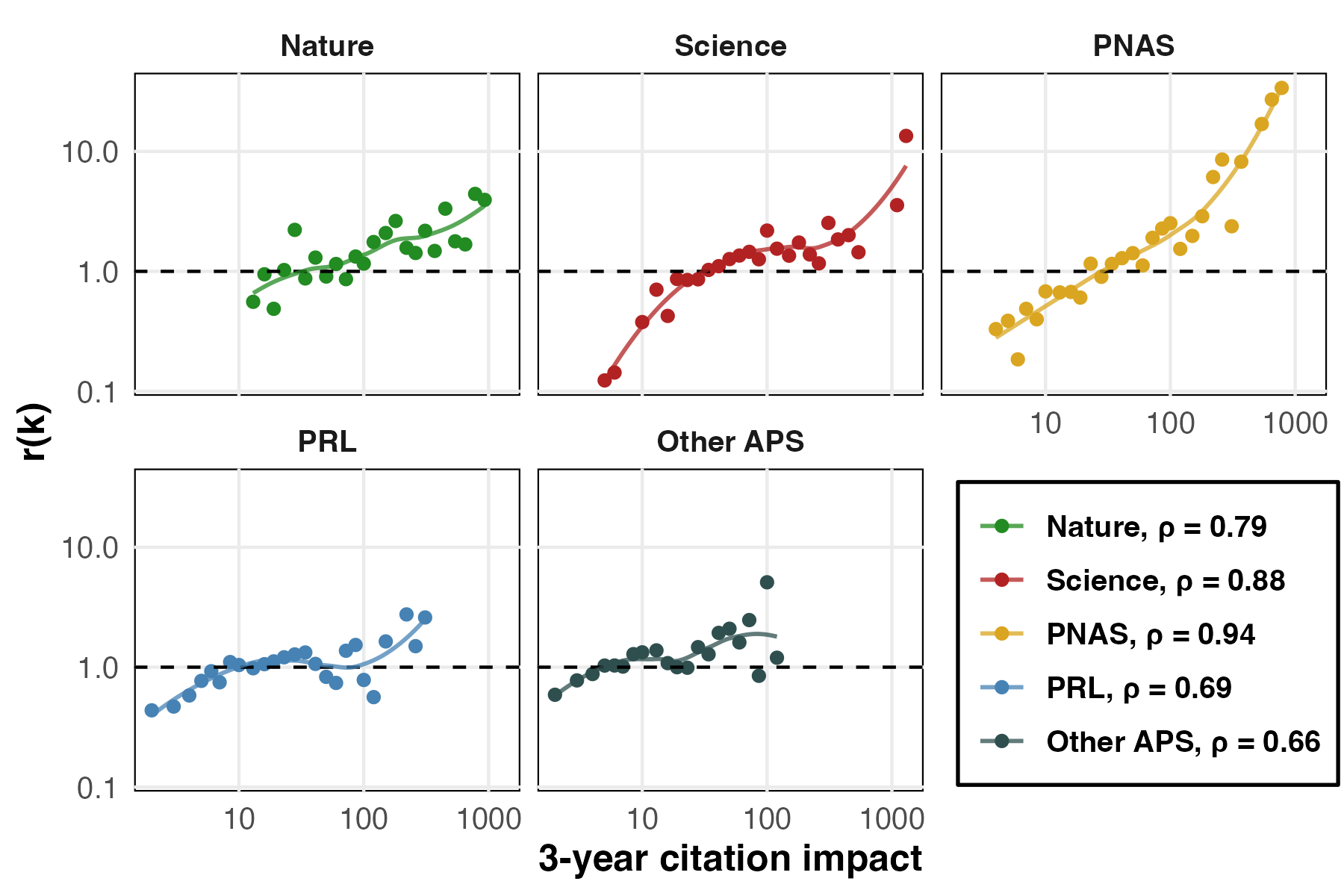}
    \caption{
    \textbf{Receipt of critical letters is associated with impact.}
    Shown is the relationship between the 3-year citation impact of papers and the likelihood of receipt of a critical letter, $r(k)$. 
    The likelihood, $r(k)$, is estimated by first diving all publications in each journal into discrete bins based on their 3-year citation impact.
    Bins are logarithmic scaled such that they grow progressively wider. 
    Then, within each bin, $r(k)$ is estimated as the ratio between the proportion of all papers targeted by a critical letter appearing in that bin and the proportion of all non-targeted papers within the same bin. 
    Each point corresponds to an estimated $r(k)$ within a particular bin and journal.
    A loess regression is shown for each journal to aid interpretation. 
    The legend shows Pearson's $\rho$ to summarize the linear correlation between impact and $r(k)$ for each journal; to compute this we use the logarithmically-scaled right-hand edge of each bin.
    This graph illustrates that the likelihood of receiving a critical letter grows roughly linearly with papers' log-impact.
    }
    \label{si:fig:impact-likelihood-scatter}
\end{figure}

Our findings align with previous research: critical letters are disproportionately directed at the most-cited papers in each journal (Fig.~\ref*{fig:paper-features}.A1 - A5).
The measure $\mu_{rank}$ represents the mean percentile rank of criticized papers' citation impact relative to all papers in the same journal. 
The strongest concentration is observed in \textit{PNAS} ($\mu_{\text{rank}} = 67.7\%$, Fig.~\ref*{fig:paper-features}.A3), and the weakest in PRL ($\mu_{\text{rank}} = 54.2\%$, Fig.~\ref*{fig:paper-features}.A4).
We also conduct one-sample Kolmogorov–Smirnov (1s KS) statistical tests comparing the percentile distribution of critical letters' impact against a uniform distribution expected if letters were randomly targeted, showing statistical significance for all journals.
The relationship between (log-scaled) citation impact and the likelihood of receiving a critical letter is roughly linear in all journals (Fig.~\ref*{si:fig:impact-likelihood-scatter}).

To further assess how well the citations alone explain the likelihood of the receipt of critical letter, we fit logistic models for each journal and using the Nagelkerke $R^2$ goodness of fit (Table~\ref*{tab:psuedor2}), find the best fit for \textit{PNAS} ($R^2 = 0.183$), and the lowest for \textit{PRL} ($R^2 = 0.030$) and \textit{Other APS} ($R^2 = 0.013$); 
field-normalized measures of impact tend to offer a slightly improved fit.
Together, these findings demonstrate the prominent role of attention in the likelihood that a paper is targeted by a critical letter.

Beyond citation impact, other paper-level characteristics may correlate with an increased likelihood of criticism.
One such feature is a paper's interdisciplinarity.
Papers that engage with or are accessible to multiple disciplinary communities may inadvertently breach the theoretical, methodological, or interpretive norms of any one of those fields.
Such transgressions invite criticism from disciplinary specialists engaged in boundary work~\cite{Gieryn1999, Klein2021} and may expose the paper to intensified scrutiny from differing disciplinary perspectives, as observed in peer review~\cite{mcleish_2016_interdisciplinary, brohman_2016_interdisciplinary}. 
We quantify a paper’s interdisciplinarity using a bibliometric measure based on the balance and diversity of fields represented among its references.
Specifically, we use Simpson's Diversity Index of the fields represented among a paper's references~\cite{chen_2015_interdisciplinary} (see \textit{\methods}), which we here call \textit{reference diversity}.
Generally, the greater the number of fields represented among a paper's references and the more evenly balanced their frequency, the greater its interdisciplinarity.

We observe that papers targeted by critical letters are concentrated among the most interdisciplinary in each journal, though to a lesser extent than observed for impact (Fig.~\ref*{fig:paper-features}.B1 - B5).
The strongest concentration is observed for \textit{PNAS} ($\mu_{\text{rank}} = 56.4\%$, Fig.~\ref*{fig:paper-features}.B3), and weakest for \textit{PRL} ($\mu_{\text{rank}} = 51.4\%$, Fig.~\ref*{fig:paper-features}.B4) and \textit{Other APS} ($\mu_{\text{rank}} = 50.3\%$, Fig.~\ref*{fig:paper-features}.B5), the last of which trends towards, but does not meet a standard of statistically significant difference from a uniform distribution based on the 1s KS test comparison. 

At least some of the difference observed for reference diversity may in fact be the result of interdisciplinarity mediating the effect between citation impact and likelihood of criticism.
That is, papers with more diverse references may tend to have higher citation impact. 
Indeed, empirical evidence demonstrates that many of the most cited papers have high levels of interdisciplinarity~\cite{chen_2015_interdisciplinary, yegros-yegros_2015_interdisciplinarity}.
To test this, we also compare each distribution against a control population of non-targeted papers matched from the same journal, with a similar year of publication and citation impact (see \methods). 
To compare we use a two-sample Kolmogorov–Smirnov (2s KS) test comparing the percentile distribution of the two populations.
The strongest effect is again observed for \textit{PNAS}, for which criticized papers are among those with the highest citation diversity ($\mu_{rank}=56.4\%$, 1s KS, $p < 0.01$, Fig.~\ref*{fig:paper-features}.B3) and distinct from the matched population (2s KS, $p < 0.01$).
For \textit{Nature}, reference diversity is distinct from the uniform distribution ($\mu_{rank}=54.1\%$, 1s KS, $p = 0.03$, Fig.~\ref*{fig:paper-features}.B1), and somewhat resembles the matched population ($p = 0.06$, Fig.~\ref*{fig:paper-features}.B1), with similar findings for \textit{Science} (Fig.~\ref*{fig:paper-features}.B2).
For \textit{PRL} criticized papers are distinct from the control population ($p < 0.01$, Fig.~\ref*{fig:paper-features}.B4).
Notably, in \textit{Other APS} journals, the matched population resembles a uniform distribution ($p = 0.31$, Fig.~\ref*{fig:paper-features}.B5), but not the matched population ($p = 0.02$, Fig.~\ref*{fig:paper-features}.B5), though interpretation is unclear considering this group includes multiple journals.
Reference diversity may, in certain contexts, be associated with criticism, but to some extent may be mediated by citation impact.

A paper may draw interdisciplinary attention even if it itself does not cross disciplinary boundaries. 
Here, we use the Simpson's Diversity Index to measure the diversity of fields represented among the papers citing a targeted paper (Fig.~\ref*{fig:paper-features}.C1 - C5).
We limit to citations accumulated by a paper within 5 years.
For \textit{Nature} ($\mu_{\text{rank}} = 54.2\%$, 1s KS, $p = 0.03$, Fig.~\ref*{fig:paper-features}.C1)  we observe evidence that targeted papers are concentrated among those with the most interdisciplinary citations.
However, this distribution is statistically indistinguishable from a control distribution of similarly-impactful papers (2s KS, $p = 0.23$), suggesting that in this journal citation diversity may play a mediating role between impact and criticism.
For \textit{Science}, meanwhile, criticized papers are not significantly concentrated among those with the highest citation diversity ($\mu_{rank}=50.8\%$, 1s KS, $p = 0.12$, Fig.~\ref*{fig:paper-features}.C2).
Only for \textit{PNAS} ($\mu_{\text{rank}} = 57.6\%$, 1s KS, $p < 0.01$, 2s KS, $p < 0.01$, Fig.~\ref*{fig:paper-features}.C3) and \textit{PRL} ($\mu_{\text{rank}} = 51.2\%$, 1s KS, $p < 0.01$, 2s KS, $p = 0.02$, Fig.~\ref*{fig:paper-features}.C4) do we observe firm evidence that the interdisciplinarity of a paper's citations is associated with receiving criticism, with the strongest effect for \textit{PNAS}. 
Notably, within \textit{Other APS} criticized papers are tend to be disproportionately concentrated among those with the least citation diversity, though they are comparable to a uniform distribution across percentile ranks ($\mu_{rank}=48.3\%$, 1s KS, $p = 0.75$, 2s KS, $p = 0.03$, Fig.~\ref*{fig:paper-features}.C5).
In general, these findings point to the limited relevance of citation diversity, with differential findings across journals, sometimes irrelevant (\textit{Science}, \textit{Other APS}), sometimes mediated by impact (\textit{Nature}, and evidence of some relevance for others (\textit{PNAS}, \textit{PRL}).

Another characteristic potentially linked to criticism is a paper's novelty.
Thomas Kuhn argued that science is firmly rooted in tradition~\cite{Kuhn1977}, suggesting that scientists often resist novel research that challenges established perspectives or disrupts prevailing orthodoxy. 
We hypothesize that highly novel papers are more likely to attract criticism. 
Empirical studies lend support to this hypothesis: progress and innovation often emerge following the deaths of influential scholars who upheld established norms~\cite{azoulay_2019_funeral};
novel work exhibits distinctive and more variable citation patterns~\cite{wang_2017_novelty};
and novel research may face additional challenges during peer review~\cite{boudreau_2016_looking,li_2017_expertise,siler_2016_review,lane_2022_conservatism,ayoubi_2021_novel}.
However, not all studies align with this conclusion---one landmark study found no penalty against novel work in journal peer review~\cite{teplitskiy_2022_novel}.

We quantify novelty using a widely recognized bibliometric measure based on the atypicality of a paper's references~\cite{uzzi_2013_atypical}.
In our results (Fig.~\ref*{fig:paper-features}.D1 - D5), we see a similar pattern across many journals.
For \textit{Nature} (Fig.~\ref*{fig:paper-features}.D1), \textit{Science} (Fig.~\ref*{fig:paper-features}.D2), \textit{PNAS} (Fig.~\ref*{fig:paper-features}.D3), and \textit{Other APS} (Fig.~\ref*{fig:paper-features}.D4), criticized papers tend to be among the most novel in their journal, but they tend to closely resemble the matched population.
This suggests novelty is mediated by citation impact---novel papers tend to have more citations, and citations are in turn associated with criticism. 
The exception to this pattern is \textit{PRL} ($\mu_{rank}=55.2\%$, 1s KS, $p < 0.01$, 2s KS, $p < 0.01$, Fig.~\ref*{fig:paper-features}.D4), which may indicate something unique about \textit{PRL} warranting further examination.
These results suggest that criticism is more frequent among novel papers, though this is often mediated by citation impact.

The likelihood of a paper receiving a critical letter may also be associated with the demographics of its authors. 
We conduct an exploratory analysis of three well-founded and methodologically tractable factors: (i) gender, inferred based on authors' first names;
(ii) seniority, defined as the years since an author's first publication; 
and (iii) affiliation prestige, operationalized based on the authors' institutions appearing within the top 30 in terms of impact from the Leiden Rankings (details in \appendix).

Evidence suggests that women may face biases in academic hiring~\cite{moss-racusin_2012_bias} and journal peer review~\cite{fox_2019_gender, hengel_2022_female}.
However, among criticized papers, we find little evidence that papers with women as first or last authors are disproportionately targeted for criticism (Fig.~\ref*{si:fig:margins_first-auhor}, Fig.~\ref*{si:fig:margins_last-auhor}).
Seniority and prestige may also influence patterns of critique. Researchers might be hesitant to criticize senior or prominent faculty, while junior faculty could be perceived as more vulnerable targets.
Our findings provide mixed evidence: senior first authors are more likely to receive criticism than junior authors within \textit{PNAS} and \textit{PRL}, but this effect does not extend to last authors. 
Additionally, no significant effects are observed based on the prestige of the authors’ institutional affiliation.
Overall, our analysis offers limited support for the role of author demographics in shaping the likelihood of receiving criticism. Modest associations are primarily tied to seniority, with no consistent patterns observed for gender or institutional prestige.

\begin{figure}[p!]
    \centering
    \includegraphics[width=1.0\linewidth]{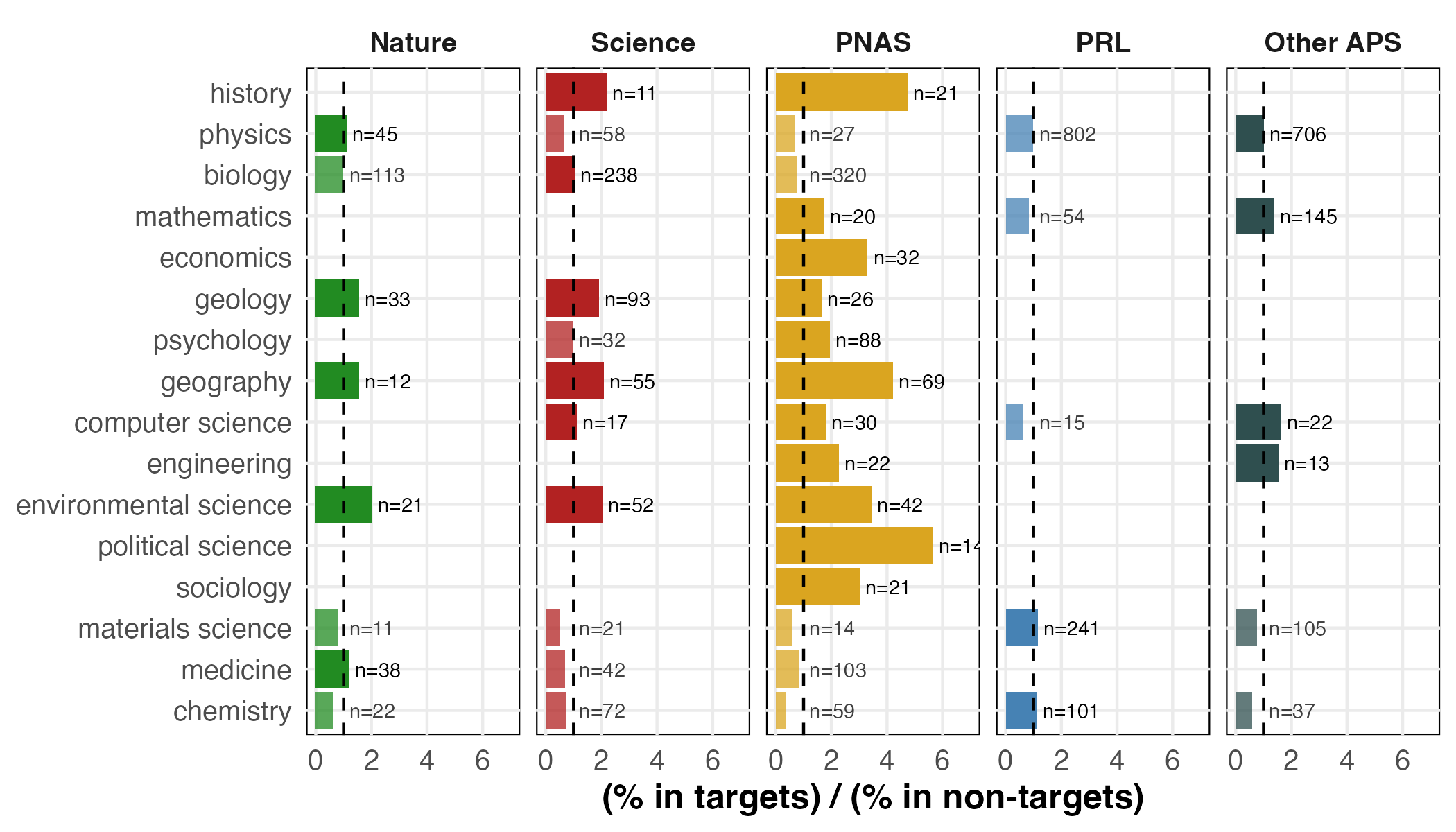}
    \caption{
    \textbf{Field representation of papers targeted by critical letters.}
    Shown for each journal is an assessment of which fields are over- and under-represented among those targeted by critical letters. 
    Bars indicate the ratio between proportion of targeted papers in each field within a journal against the proportions of non-targeted papers.
    A value greater than one indicates that targeted papers are over-represented whereas a value less than one indicates under-representation. 
    Shown for each bar is the count of papers targeted by a critical letter in that field category.
    Fields with fewer than 10 papers in a journal are not shown.
    When a paper is categorized into multiple fields it is counted for each; as such the sum of the values of $n$ will be greater than the targeted papers. 
    }
    \label{si:fig:field-representation}
\end{figure}

\begin{figure}
    \centering
    \includegraphics[width=1\linewidth]{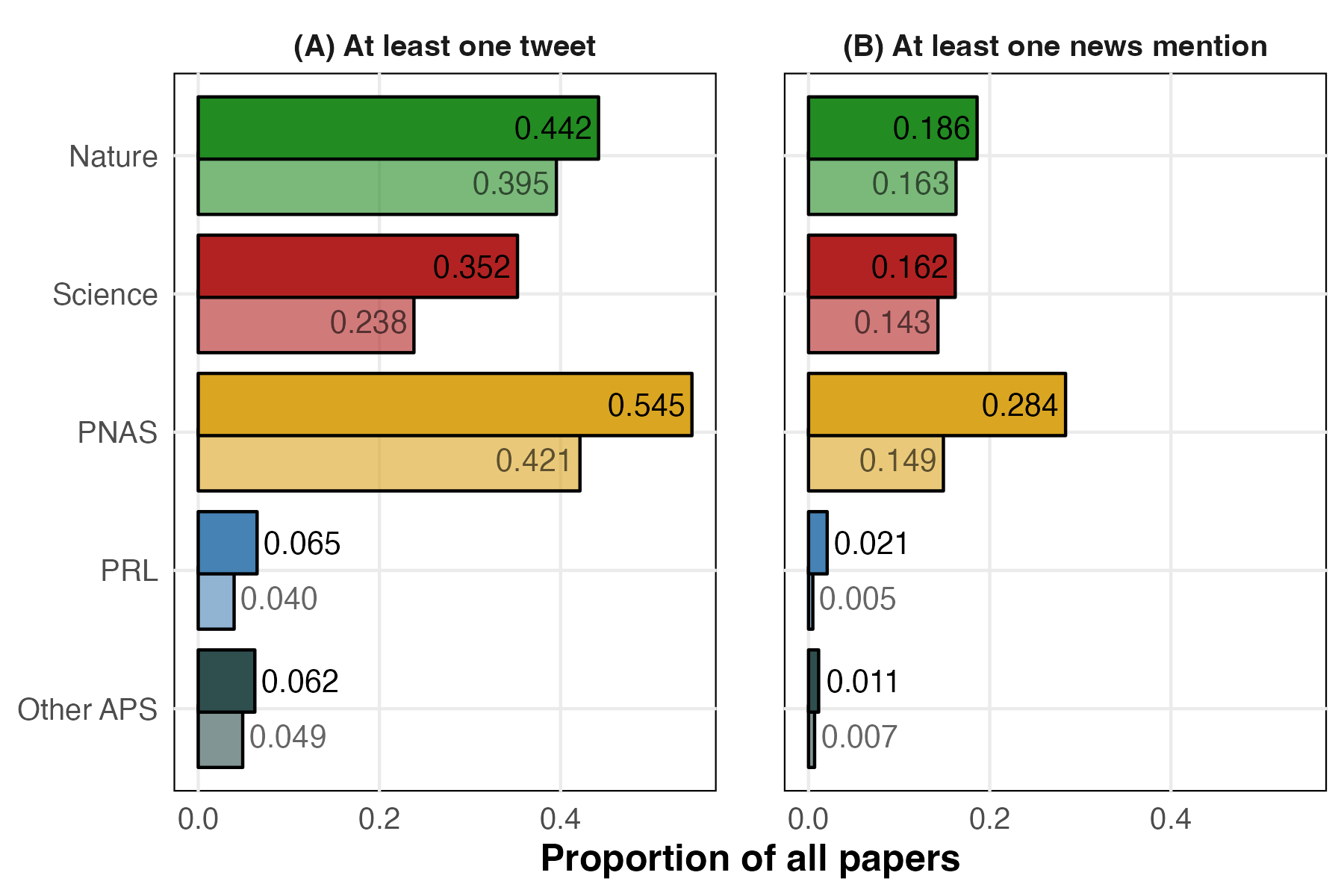}
    \caption{
        \textbf{Criticized papers more likely to have news coverage or social media engagement.}
        Shown is \textbf{(A)} the proportion of papers targeted by a critical letter that have at least one engagement on Twitter/X, and \textbf{(B)} the proportion that have received at least one mention in a news story indexed by Altmetric.com.
        Data on media engagement is sourced from SciSciNet.
        Presented here is the publication data underlying Fig~\ref*{fig:paper-features}, consisting of matched pairs of criticized papers and comparable peers.
        The proportion is shown across both the criticized papers (dark color) and the control group (light color). 
    }
    \label{si:fig:media_engagement_proportion}
\end{figure}

\begin{figure}
    \centering
    \includegraphics[width=1.0\linewidth]{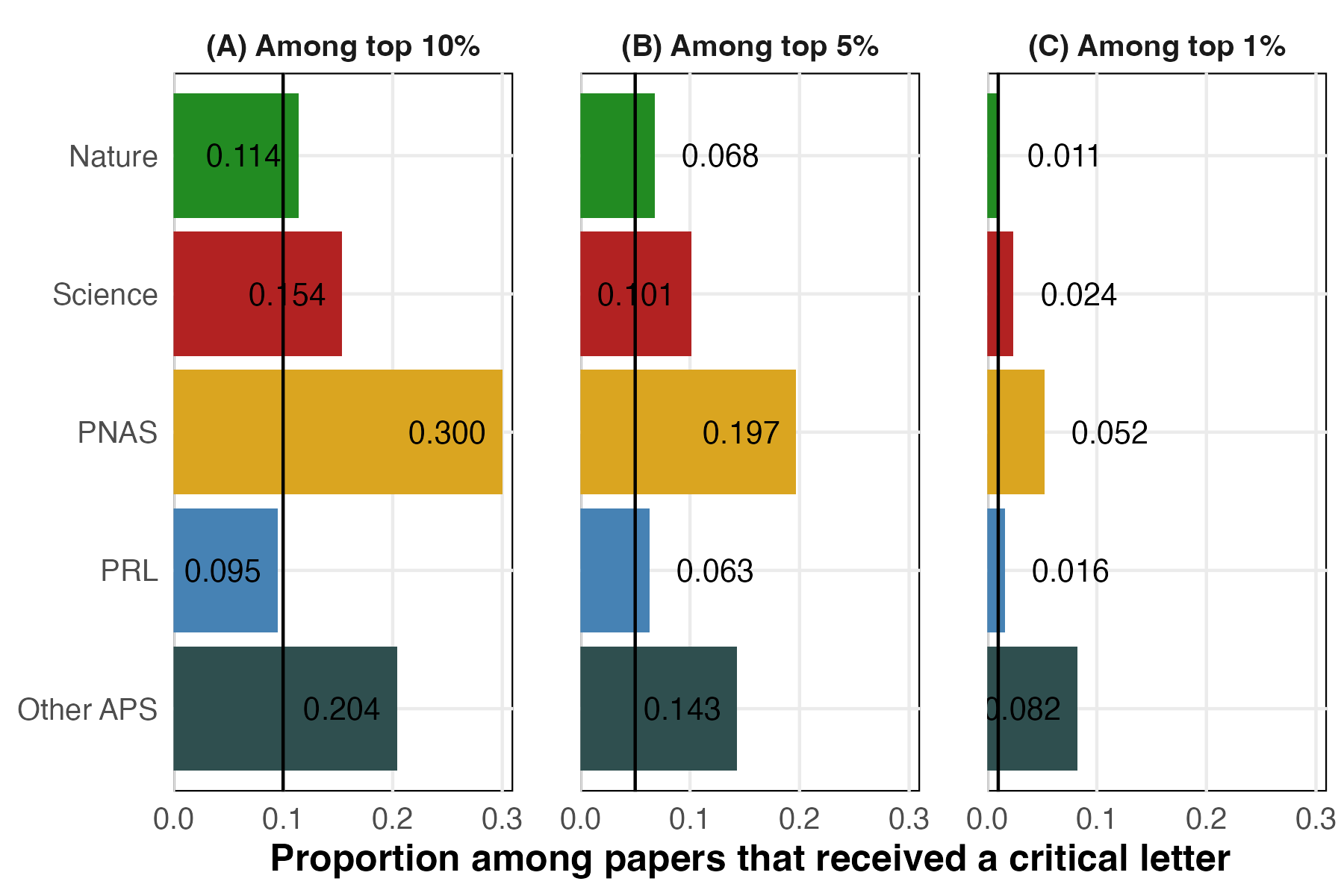}
    \caption{
    \textbf{Criticized papers are concentrated among those with the most engagement on Twitter/X.}
    Shown is the proportion of criticized papers that are within the \textbf{(A)} top 10\%, \textbf{(B)} top 5\% and \textbf{(C)} top 1\% of papers in terms of social media engagement (mentions on Twitter/X) among all papers with at least one engagement within each journal. 
    Data on social media engagement is sourced from SciSciNet.
    The black line indicates the percentile threshold for each metric (10\%, 5\%, and 1\%). 
    }
    \label{si:fig:tweet_distribution}
\end{figure}

A variety of paper-level characteristics may influence the likelihood of receiving criticism. One such characteristic is the paper’s \textit{discipline}, as some fields may be more prone to critical discourse.
We examine the representation of the disciplines of criticized papers compared to the journal as a whole (Fig.~\ref*{si:fig:field-representation}). 
Doing so, we observe that across the three multidisciplinary journals (\textit{Nature}, \textit{Science}, and \textit{PNAS}), fields such as \textit{Geology} and \textit{Environmental Sciences} are over-represented, affirming similar findings regarding the earth sciences in past work~\cite{lamers_2021_disagreement}. 
Social sciences disciplines, among them \textit{History}, \textit{Geography}, and \textit{Political Science}, are also over-represented.
We also explored characteristics that, while lacking rigorous measures or firm theoretical grounding, still warrant attention.
Prior research indicates that papers eventually retracted often receive outsized early attention on social media~\cite{peng2022retracted, serghiou_2021_attention}. 
Similarly, we find that papers receiving critical letters are more likely to have been mentioned in at least one tweet (now X post) or news story (Fig.~\ref*{si:fig:media_engagement_proportion}).
Moreover, criticized papers are disproportionately concentrated among those with the highest levels of social media engagement (Fig.~\ref*{si:fig:tweet_distribution}).
Although exploratory, these findings suggest a connection between media exposure and criticism. Media attention may broaden a paper’s readership to a larger, potentially more critical audience, or alternatively, provocative topics may simultaneously attract both media attention and critique.

In summary, across all journals, papers receiving criticism are disproportionately high-impact, with the strongest associations observed in elite multidisciplinary journals. These papers also exhibit higher levels of interdisciplinarity and novelty, although these effects are generally weaker, context-dependent, and often mediated by impact.

\subsection*{The consequences of criticism}
The effect of criticism on the citation impact of a paper remains an open question. Competing hypotheses propose that a critical letter could either increase or decrease the target paper's subsequent citations.
On one hand, critical letters may act as a kind of negative assessment highlighting methodological or interpretative flaws that may introduce doubt and thereby reduce the paper’s scholarly impact. 
Under this view, a critical letter functions similarly to a retraction, albeit the consequences are likely to be less severe~\cite{peng2022retracted,azoulay_2015_retractions}.
On the other hand, criticism may, regardless of its content, amplify the paper's visibility, consistent with the adage that ``there is no such thing as bad publicity''~\cite{berger_2010_publicity,radicchi_2012_comment}. 
According to this view, the critical letter could call newfound attention to its target, potentially counteracting a natural decline in citations over time.
A third possibility is that the critical letter has no effect on the paper’s citations.
This could occur if the letter is not widely read, or if it is read but ultimately dismissed as inconsequential, perhaps due to a compelling rebuttal by the original authors or a perception that the criticism lacks substantive impact on the paper’s value.

We examine the effects of critical letters on citation trajectories through a quasi-experimental design inspired by previous studies of retractions~\cite{peng2022retracted}.
Specifically, we compare the citation patterns of papers that received critical letters (treatment group) to those of a matched set of similar papers that did not receive criticism (control group).
Control papers were selected based on publication in the same journal, at a similar time, within the same field, and with similar citation impact at the time the critical letter was issued (see \methods).
Across all journals, a large majority of criticized papers were matched with a comparable control (Table~\ref*{si:tab:paper-matched-counts}). 
This matching procedure yielded high-quality comparisons, as evidenced by the close alignment in impact metrics between treatment and control papers (Fig.~\ref*{si:fig:diagnostics}).
Our outcome measure is the growth in cumulative citations over the three years following the critical letter, relative to citation counts prior to the criticism. 
By comparing citation growth within each paper, we better account for sources of heterogeneity not captured by matching, such as fine-grained research topics.

\begin{figure}
    \centering
    \includegraphics[width=1.0\linewidth]{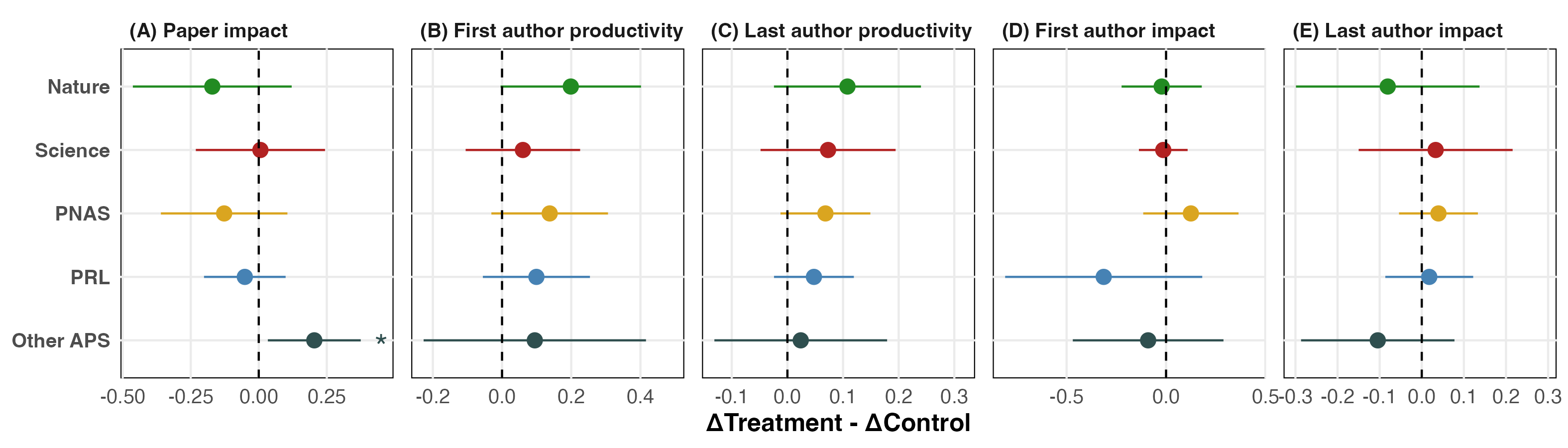}
    \caption{
    \textbf{The consequences of criticism.}
    The x-axis shows, from left to right:
    \textbf{(A)} Comparison between the cumulative impact growth of papers targeted by a critical letter compared to a matched population of control papers from the same journal, within one year of publication, the same field, and with similar impact. 
    Here,  change in ``paper impact'' ($\Delta$Treatment and $\Delta$Control) is defined as the ratio between the citations received by a paper after receipt of a critical letter compared to those received prior; in the case of the control population we use an equivalent time lag.
    \textbf{(B-C)} Comparison of the change in the average yearly fractional productivity before and after the critical letter for first and last authors of targeted papers compared against a control population of authors sampled from the same journal and with similar prior performance. 
    \textbf{(D-E)} Comparison of the change in average impact prior for first and last authors of targeted papers compared against the control population.
    Here, author impact refers to the ratio in average field-normalized 3-year citation impact of papers published in the five years before and after receipt of the critical letter. 
    For matching papers, citation tolerance is set to 5\%. 
    For matching authors, both citation and productivity tolerance is set to 10\%.
    Error bars correspond to 95\% confidence intervals.
    For each comparison we conduct paired t-tests comparing the treatment and control groups, with asterisks included to indicate the significance level (``*'' when $p < 0.05$); these tests are to guide interpretation, and not for confirmatory analysis.
    }
    \label{fig:matched-results-impact}
\end{figure}

\begin{figure}[p!]
    \centering
    \includegraphics[width=1.0\linewidth]{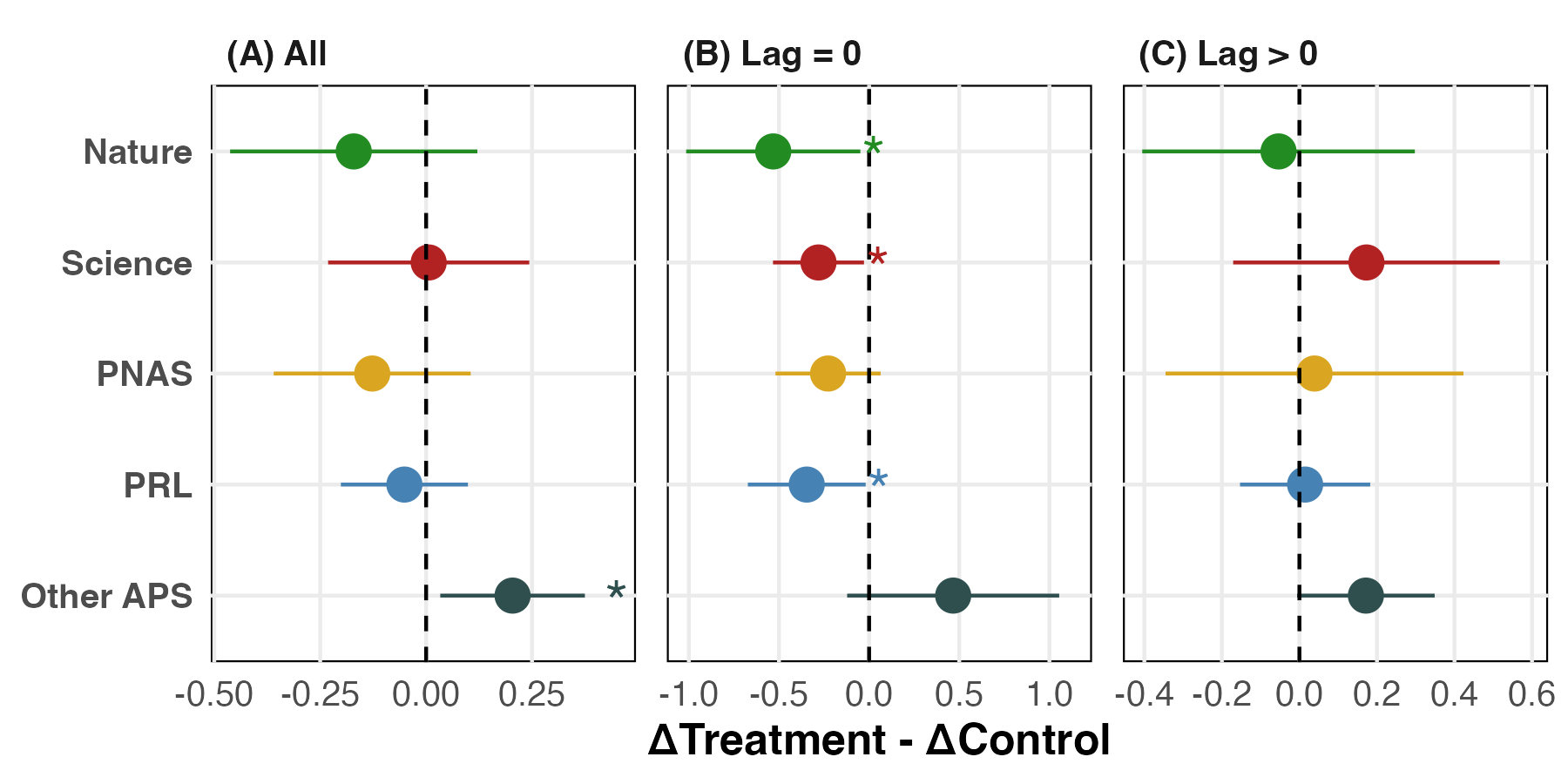}
    \caption{
    \textbf{Comparison of the matched papers against matched control by time lag.}
    The x-axis shows a comparison between the cumulative impact growth of papers targeted by a critical letter compared to a matched population of control papers from the same journal, time period, field, and with similar impact. 
    Here,  change in ``paper impact'' ($\Delta$Treatment and $\Delta$Control) is defined as the ratio between the citations received by a paper after receipt of a critical letter compared to those received prior; in the case of the control population we use an equivalent time lag.
    \textbf{(A)} shows the entire data, which is the same as in Fig.~\ref*{fig:matched-results-impact}.
    \textbf{(B)} shows the subset of papers in which the critical letter is published in the same year as its target.
    \textbf{(C)} shows the subset of papers in which the critical letter was published one or more years following its target. 
    }
    \label{si:fig:consequence_time-lag}
\end{figure}

We find no statistically significant difference in citation growth between criticized papers and their matched peers in the control group across four of the journals analyzed (Fig.~\ref*{fig:matched-results-impact}.A).
The only exception is for \textit{Other APS} journals ($t = -0.2, p = 0.020$) in which results suggest a slight tendency for criticized papers to receive more citations than the control.
Whereas these results are based on the most restrictive matching criteria, we also examine more lenient parameters (Tables \ref*{si:tab:paper-matched-ttest}); most combinations result in a small observed effect for the \textit{Other APS} journal, but these results only just meet the conventional $p < 0.05$ threshold and would not withstand stricter significance criteria or post-hoc corrections for multiple comparisons.
The timing of the critical letter relative to the original paper's publication may also play a role. 
Notably, papers receiving a critical letter within the same year of publication tended to show \textit{lower} than expected citation growth in \textit{Nature}, \textit{Science}, and \textit{PRL}, suggesting that immediate criticism might have a dampening effect on citations in this specialized journal (Fig.~\ref*{si:fig:consequence_time-lag}); whether this is the result of noise or an actual effect of criticism in \textit{PRL} deserves examination in a future study.
Overall, these findings indicate minimal or negligible impact of critical letters on the citation trajectories of the targeted papers, with the strongest effect observed for the \textit{Other APS} journal, and other small observed effects likely attributable to statistical noise.

While critical letters appear to have limited consequence for the citation trajectories of targeted papers, authors of criticized papers may still be concerned about potential career repercussions.
As with papers, plausible arguments can be made for both positive and negative effects of criticism on career prospects.
On one hand, receiving criticism could be perceived negatively by the broader scientific community, potentially stigmatizing the author and impairing their capacity to secure resources or carry out high-quality research, similar to the ``scandal'' consequences of retractions~\cite{azoulay_2017_scandal}.
On the other hand, the act of defending one's work against criticism may stimulate new research ideas, offering creative and intellectual benefits that catalyze new research ideas that lead to more impactful publications.

To investigate these potential career effects, we employ a similar quasi-experimental designed to use for papers.
For each lead author (first or last authors) of a paper targeted by a critical letter (treatment group), we identify a set of comparable peer authors as a control group.
These control authors are selected among those who published in the same journal around the same time, with the same authorship role (first or last), similar average impact and productivity over the preceding five years, and a comparable career age (see \methods).
As with papers, outcome measures are defined for each researcher as the growth in metrics that capture their average productivity or impact;
growth is calculated between two periods: the five years preceding receipt of the critical letter, and the five years after.
The difficulty in matching authors means that far fewer end up matched as compared to papers, fewer first authors than last authors, and varies greatly depending on the choice of matching parameter (Table~\ref*{si:tab:authors-matched-counts});
we highlight results using a criterion of 10\% impact and productivity tolerance that balances quality and quantity of matches, though analysis is repeated across all combinations.
We find that while we lose many candidates, the resulting matches are high quality with close alignment in terms of impact and productivity prior to receipt of the critical letter (Tables~\ref*{si:tab:first-author-match-quality},\ref*{si:tab:last-author-match-quality}).

We first consider authors' productivity, defined as the sum of their fractionalized authorship contributions in the period before and after receiving a critical letter. 
Fractionalized authorship is calculated as their fractional contribution to the authorship of a paper, or $\frac{1}{\text{\# authors}}$.
We assume that authors' productivity naturally changes over the course of their career, and for each author calculate the ratio of their productivity prior to and post criticism;
by keeping comparisons within authors, we also ameliorate the effects of characteristics not controlled for during matching. 
Our results demonstrate no statistically significant difference in productivity between the treatment and control groups for first authors (Fig.~\ref*{fig:matched-results-impact}.B) or last authors (Fig.~\ref*{fig:matched-results-impact}.C). 
We repeat the matching process with more lenient parameters (Table~\ref*{si:tab:authors-matched-impact-ttest} and while there are exceptions, overall they provide no strong evidence for any consequences of criticism on the productivity of the authors of the targeted paper. 

We apply a similar method to evaluate how criticism influences the impact of an author's papers. 
Author impact is quantified as the average field-normalized 3-year citation impact of papers published in the five years after the critical letter, relative to those from the preceding five years.
As with productivity, we account for career-stage variations, we calculate the ratio of impact between these periods for each author.
Across all journals we observe no significant difference between the average impact growth before and after the criticism between treatment and control groups for both first (Fig.~\ref*{fig:matched-results-impact}.D) and last authors (Fig.~\ref*{fig:matched-results-impact}.E).
This null result remains consistent across various combinations of matching parameters (Table~\ref*{si:tab:authors-matched-impact-ttest}).

In summary, our analysis is consistent with the conclusion that critical letters are inconsequential for the citations accumulated by papers they target as well as the productivity and impact of their authors.

\subsection*{The visibility of criticism}

Given the emphasis on debate and disagreement in theories and histories of scientific progress, it is surprising that, even in its most direct form---the critical letter---criticism appears to have no consequence for the citation trajectories of targeted papers and authors.

\begin{figure}[h!]
    \centering
    \includegraphics[width=1.0\linewidth]{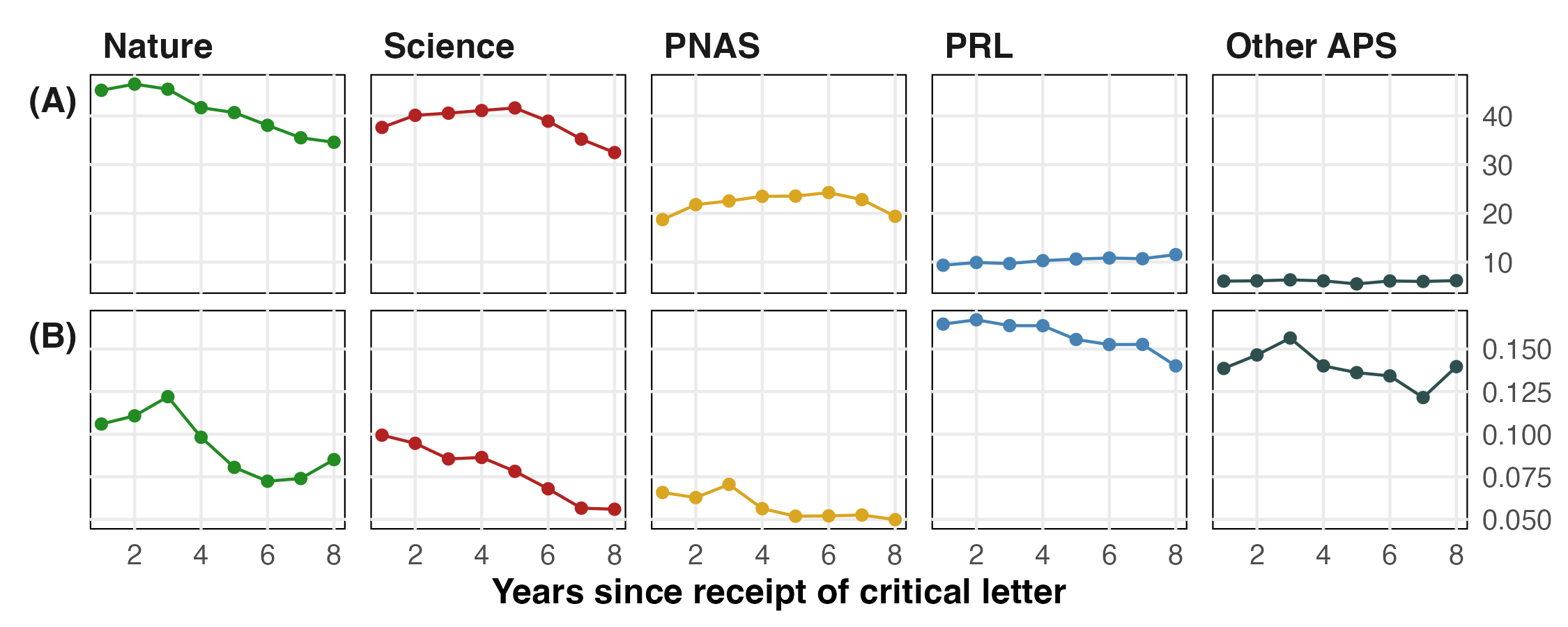}
    \caption{
    \textbf{Critical letters have low impact despite continued citation to criticized papers.}
    The x-axis shows time normalized as the number of years since receipt of a critical letter. 
    The y-axis shows:
    \textbf{(A)} the pooled average number of citations received each year by papers that were targeted by a critical letter;
    \textbf{(B)} the pooled average percentage of citations to the targeted paper which also cite the critical letter.
    Author self-citations, defined as any overlap between authors of the cited and citing paper, are excluded in all plots.
    Replies to critical letters are also excluded.
    }.
    \label{fig:visibility}
\end{figure}

One explanation is that criticized papers receive fewer citations than the counterfactual scenario in which no criticism was issued.
That is, other researchers who may potentially cite the paper read the criticism and opt not to cite, but this effect is masked by the large impact of the target paper. 
Without data on which researchers were exposed to a critical letter, we are unable to test this explanation directly. 
However, available evidence suggests that this explanation is unlikely.
Our analysis provided no evidence that papers have fewer citations than expected after being criticized (Fig.~\ref*{fig:matched-results-impact}). 
We also see that after receiving criticism, targeted papers continue to receive a high number of citations (Fig.~\ref*{fig:visibility}) and follow steady trends.

A second explanation is that researchers are exposed to the critical letter but do not change their citing behavior.
They may find the criticisms unpersuasive or because the citation serves a purpose unrelated to the validity of the criticized work.
In this scenario, one might expect researchers citing the original paper to also cite the critical letter for context.
However, we observe that only a small fraction of citations to the targeted paper are accompanied by a citation to the critical letter—around 10\% for elite generalist journals and 15\% for APS journals in the initial years, and declining over time (Fig.~\ref*{fig:visibility}).

\begin{figure}
    \centering
    \includegraphics[width=1\linewidth]{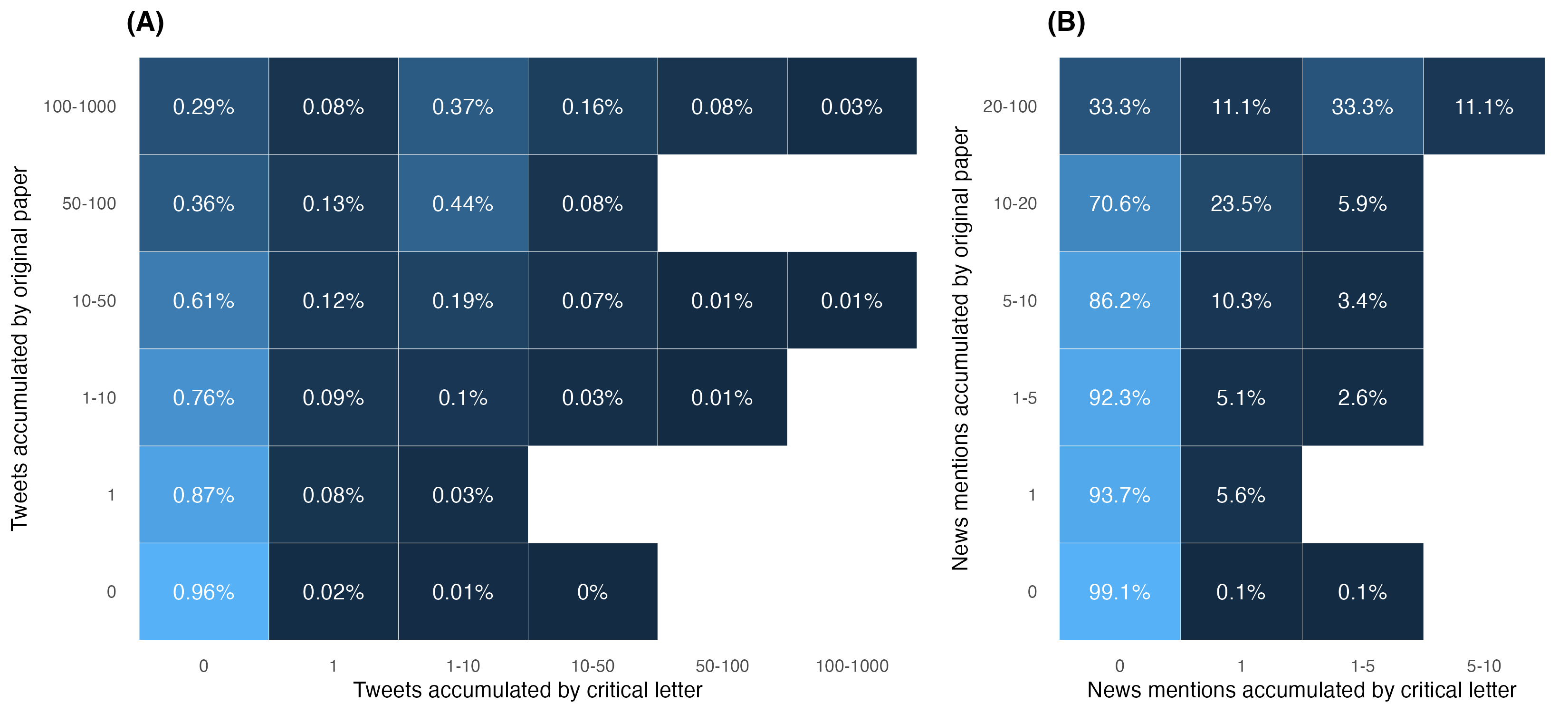}
    \caption{
    \textbf{Altmetrics of critical letters compared to their target.}
    On the x-axis is \textbf{(A)} the number of tweets and \textbf{(B)} number of mentions in news articles received by critical letters, discretized into ranges.
    The y-axis shows the same ranges for papers targeted by each critical letters.
    Each cell shows the number of occurence in each pair of bins as a proportion across each row.
    For example, for original papers that received exactly 1 tweet, 90\% of their paired critical letters received zero tweets, and 6\% received exactly 1 tweet. 
    }
    \label{si:fig:letter_altmetrics}
\end{figure}

While we assume that researchers aware of a critical letter would likely co-cite it with the original paper, we cannot confirm this assumption without data on researchers' exposure to specific letters.
However, we argue that a lack of visibility is a more likely explanation. 
This view is supported by altmetric counts sourced from \textit{SciSciNet}, which show that critical letters receive considerably fewer tweets and news mentions than the papers they critique (Fig.~\ref*{si:fig:letter_altmetrics}).
Assuming that altmetric measures serve as reasonable proxies for visibility, these findings suggest that critical letters reach a much smaller audience than the original papers, potentially limiting their influence on broader citation behaviors.

\begin{figure}
    \centering
    \includegraphics[width=1\linewidth]{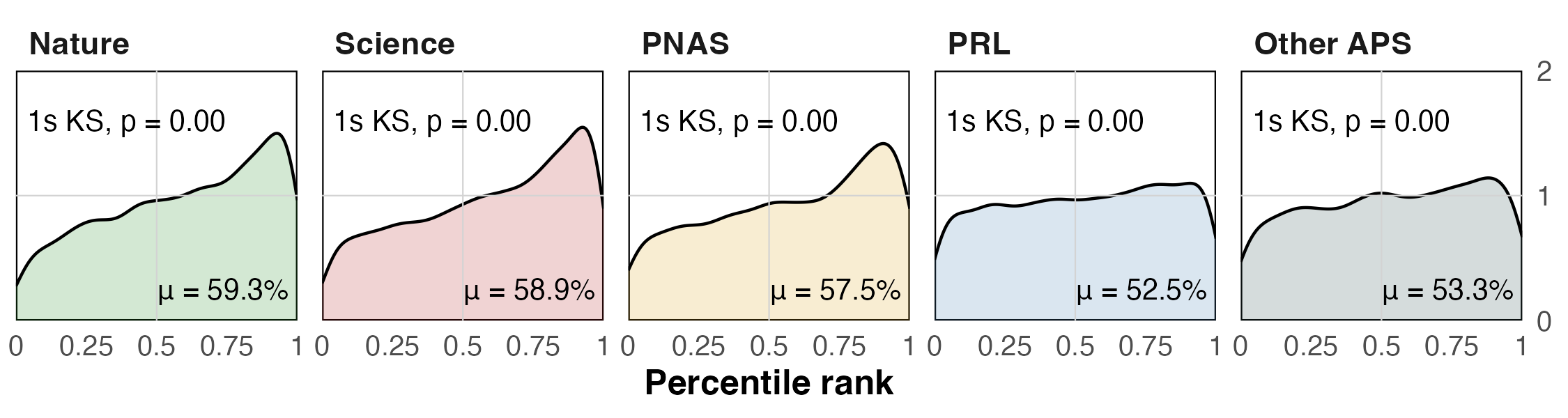}
    \caption{
    \textbf{Papers similar to the criticized study co-cite the critical letter}.
    Shown are the distribution of similarity ranks of studies that cite both the study which received a critical letter along with the letter itself.
    Similarity ranks are computed for each criticized paper and based on the population of all citing papers;
    for example, a rank of 0.9 indicates that the citing paper is more similar to the criticized study than 90\% of its citations.
    For each journal, $\mu$ refers to the average percentile rank of co-citing paper similarity.
    We conduct two \textit{Kolmogorov–Smirnov} tests to guide interpretation. “1s KS” refers to a one-sample, one-sided Kolmogorov–Smirnov test comparing the distribution of percentile ranks of co-citing papers against a uniform distribution. 
    low p-value (by convention, $p < 0.05$) suggests that co-citing papers are concentrated among the most similar studies.
    Results are shown in greater detail in Table~\ref*{si:table:embedding-tests}.
    }
    \label{fig:embedding}
\end{figure}

The visibility of critical letters is not uniform across all readers.
Those authors most familiar with the field of the criticized paper are most likely to encounter the letter and consider it relevant to cite, whereas those readers from more distant fields may see the original paper but not its criticism. 
To test this, we generate vector representations for each paper targeted by a critical letter as well as for all papers that cite it (see \methods).
The cosine distance between these vectors provides an approximate measure of topical similarity.
A similar approach has been used previously to evaluate the topical distance between retracted papers and their post-retraction citations~\cite{woo_2024_retractions}.
After excluding replies and self-citations, we find that papers co-citing the critical letter tend to be among the most topically similar to the targeted paper (Fig.~\ref*{fig:embedding}).
This effect is most pronounced in elite generalist journals such as \textit{Nature} ($\mu = 59.3$), \textit{Science} ($\mu = 58.9$), and \textit{PNAS} ($\mu = 57.5$).
For the specialist \textit{APS} journals, the effect is present but weaker, possibly because these journals primarily attract citations from within their own field, reducing the likelihood of cross-disciplinary citations.

In summary, critical letters appear to have limited visibility. 
Our findings suggest that critical letters are most visible within their immediate disciplinary communities, whereas researchers in more distant fields may be less likely to encounter or cite them.
This pattern implies that the influence of critical letters may be constrained by disciplinary boundaries, with limited impact on the broader academic audience.

\section*{Discussion}
Criticism is widely regarded as a fundamental component of science, central to theories of scientific progress and pervasive throughout its history.
Critical letters represent one of the clearest and most explicit forms of criticism in science, making them an effective case study for examining the origins, consequences, and visibility of criticism in science.

This study highlights the central role of attention in the origins of criticism: as a paper receives more citations (or online engagement), it is more likely to have been targeted by a critical letter. 
One possible explanation for this link is that attention correlates with certain paper-level characteristics that invite critique. For instance, papers that are more provocative or more prone to errors may be especially likely to elicit criticism. Indeed, our analysis finds that papers targeted by criticism are disproportionately interdisciplinary and novel. However, these associations are relatively weak, vary across journals, and may be mediated by impact.
Another explanation is editorial policy.
Upon receiving a critical letter, editors decide whether to publish it, potentially factoring in the popularity of the target paper; yet, because we lack data on rejected submissions, we cannot fully disentangle the effects of editorial policy.
Nonetheless, the consistent association between attention and criticism across journals with different editors and policies suggests a non-editorial explanation.
A third explanation is that attention itself invites scrutiny. This has been previously proposed to explain why retractions are disproportionately common in large, generalist journals~\cite{cokol_2007_retracted}.
According to this view, the likelihood of receiving criticism is generally uniform across papers within a journal: the more a paper is read, the higher the chance that a reader will identify an issue and be motivated to submit a letter.
Under this explanation, interdisciplinarity, novelty, and other paper-level characteristics influence criticism primarily by increasing attention.
We argue that this explanation is not only the most intuitive but also the one most consistent with our findings.

Regarding the consequences of criticism, our study finds no strong evidence of any effect---neither on the citation impact of papers targeted by criticism nor the future performance of their authors.
Our results align with an epiphenomenal view of criticism: it appears to be merely a byproduct of attention, and does not bring newfound attention nor detract from the attention that a paper receives.
This finding is surprising, as criticism plays a critical role in theories of scientific progress and should be expected to have some consequence, whether positive or negative, on the papers and authors involved. 
Indeed, contradictory replications have been shown to prompt scientists to revise their beliefs about an original study~\cite{mcdiarmid_2021_replication}; one would expect a similar adjustment in response to criticism, and yet none are observed. 

One possible explanation for the absence of impact is that critical letters may lack substance.
That is, they may not offer critiques of sufficient merit or relevance to influence readers' citing behaviors.
However, we believe this explanation to be unlikely:
a manual review of critical letters indicates that most are backed by substantial efforts, such as replications or new data (see \appendix). Notably, letters in \textit{PNAS} less frequently feature novel work, and yet yield results similar to those in \textit{Nature} and \textit{Science}, suggesting that, at least among the journals we study, the substance of criticisms is unrelated to their consequences.
It is also possible that replies consistently address the critiques.
However, replies tend to have lower impact and receive less attention than critical letters, suggesting that they have even less visibility (Table~\ref*{si:tab:letters-and-replies-visibility}).
Taken together, this evidence suggests that critical letters are indeed substantive, indicating the need for an alternative explanation.

When it comes to the consequences of criticism, our study finds no strong evidence for any, neither for the impact of papers targeted by criticism nor for the productivity and impact of their authors. 
Our results are consistent with an epiphenomenal view of criticism as a byproduct of attention but with no causal effect on the attention that the paper receives. 
This result is surprising.
Criticism occupies a fundamental role in theories of scientific progress.
Yet when it comes to what is perhaps the most explicit and formalized form of criticism we observe no evidence of change in citing behavior or author impact. 
Indeed, when exposed to contradictory replications, scientists have been observed to update their beliefs about the original study~\cite{mcdiarmid_2021_replication}; we would expect a similar update for criticism.
One explanation is that the critical letters are not actually substantive. 
That is, their critiques lack in merit or significance, and so are not pertinent to the decision to cite a paper.
The sum of evidence makes this explanation unlikely. 
Our manual annotation of critical letters finds that the majority are often backed by substantial efforts such as partial replication and the introduction of new data (see \appendix).
Notably, letters in \textit{PNAS} have the lowest incidence of novel work, and yet results are virtually identical to those in \textit{Nature} and \textit{Science}, suggesting that the level of evidence has no relationship to the consequence of criticism. 
It may be that in all cases the replies of the original authors adequately respond to critiques.
However, this explanation is unlikely as replies have less visibility than even the critical letters they are rebutting (Table~\ref*{si:tab:letters-and-replies-visibility}).
Taken together, this evidence suggests that critical letters are indeed substantive, or at least that their substance is not relevant to our findings.

We propose that the lack of impact of critical letters may be due to their limited visibility.
Specifically, while authors read the original paper, they often overlook the associated critical letter, having had no exposure to it.
Our results support this view, revealing that critical letters receive only a fraction of the citations of the papers they target, even as the target papers continues to accumulate citations years after its publication. 

Why are these critical letters overlooked?
When searching the literature, researchers tend to follow one of two strategies~\cite{woo_2024_retractions}.
The first is an \textit{engaged} approach, in which authors actively interact with the literature and cite only what they have read thoroughly.
The second is a \textit{heuristic} approach, where citations are made without fully reading or understanding the paper.
Heuristic approaches may involve lifting citations directly from the references of another paper, incorporating citations used in one's previous work without knowledge of ongoing developments, or referencing work based only on cursory readings. 
Many case studies highlight such heuristic processes in relevant contexts.
Retracted studies and misquoted findings often persist in the academic literature due to lack awareness of the retraction~\cite{schneider_2020_retraction, leung_2017_letter, hsiao_2021_retracted}.
In one notable case, a critical article was frequently cited in support of the very article it criticized, indicating that even citations to the critical letter may sometimes lack genuine engagement~\cite{stang_2018_quotation}.
Thus, authors following a heuristic approach to citation may cite the original paper without realizing that a critique exists.

According to this interpretation, researchers who follow an \textit{engaged} citation approach are more likely to be aware of and cite critical letters.
Researchers are particularly inclined to engage in this way when citing articles within their own field.
Our findings support this notion: papers that co-cite both the critical letter and the targeted article are typically the most topically similar to the target.
Together, these observations suggest that authors aware of critical letters tend to cite them, while most citations to the targeted paper likely occur without full awareness of the critiques it has received.

Visibility also provides a useful perspective for understanding why our findings vary across journals.
Criticism is most strongly associated with impact in elite generalist journals, while this association is weakest in \textit{PRL} and other \textit{APS} journals.
Similarly, the link between citing paper similarity and co-citation of the critical letter is less pronounced in these journals.
Journals like \textit{Nature}, \textit{Science}, and \textit{PRL} attract broad readerships, and papers published there are likely to be seen and cited by authors from distant fields who may not be deeply engaged with the specific literature.
In contrast, \textit{APS} journals focus on physics and reach a narrower readership, such that readers are more likely to share a similar disciplinary context with the criticized paper and may employ more engaged citation practices.

This study should be understood in the context of its limitations.
First, our focus on critical letters in a select group of journals restricts the generalizability of our findings to other scientific fields and broader forms of scientific disagreement;
one goal of our work is to establish a framework that can be applied to other contexts. 
Second, while we attempt to isolate the effects of critical letters, there remain unobserved confounding factors such as a study's rigor and provocativeness, among others, that might shape the likelihood of receiving criticism or its consequences. 
Additionally, there are differences in editorial policy between journals; in \textit{PNAS} for instance, critical letters must be submitted to the journal within six months of the publication of its target;
allowable word count, number of figures, and number of references also vary across journals.
Third, while we argue that critical letters are the most explicit and formal instances of criticism in science, critique is ubiquitous and is likely to occur elsewhere: within original research articles, outside scholarly literature via blogs, social media, or private correspondence.
Fourth, we focus on the effects of criticism as they manifest in papers' citations and authors' performance; 
this provides only a limited view of the potential consequences of criticism that future work could expand upon.

We emphasize that our findings are descriptive rather than normative.
While the limited impact of criticism is surprising, the question of whether criticism should influence a paper's impact or an author’s performance is one for the broader scientific community to address. 
Our aim is to assure that our findings provide an empirical foundation for future efforts to align the role of criticism with the norms and ideals of science. 

What our findings do provide is a useful framework for understanding another form of criticism in science: post-publication peer review.
This alternative form of peer review has gained increased popularity as a potential remedy for many of the limitations of traditional pre-publication peer review.
Specifically, it is believed that post-publication peer review will enable the rapid dissemination of findings and transparent evaluation by the broader scientific community~\cite{bastian2014stronger, winker2015promise}.
Platforms such as \textit{F1000} and \textit{PubPeer} have made strides in popularizing post-publication peer review~\cite{thelwall_2020_open, ortega2021pubpeer}.
``Red Teaming'', as is used in the tech sector to probe security vulnerabilities, has also been proposed as a formal practice to unearth errors in published work~\cite{lakens_2020_critics}.
However, because this practice remains relatively new and limited in scale, it is not yet clear how effectively it will function when implemented more broadly.
Critical letters provide a natural analogue for studying post-publication peer review and may even represent a best-case scenario.
Publishing a critical letter is challenging, often constrained by tight timelines, word limits, and editorial discretion;
as a result, the critiques published after this gauntlet are likely to be rigorous and substantive.
However, our findings reveal significant challenges for the broader adoption of post-publication peer review.
Like critical letters, post-publication peer review may end up disproportionately focused on high-impact papers and suffer from limited visibility.
We believe these to be problems that post-publication peer review will need to address if it is to be implemented on a wide scale. 

Our findings further illustrate a recurring pattern in science: while initial claims garner significant attention, their subsequent evaluations receive only a fraction of that attention. 
This phenomenon has been documented in the context of retractions~\cite{woo_2024_retractions, peng2022retracted}, large-scale replication studies~\cite{hardwicke_2021_contradictory, schafmeister_2021_replication}, and now with our study of post-publication critiques. 
These practices represent some of the few formal mechanisms available for reconciling competing truth claims in science. However, each---albeit to varying degrees---demonstrates limitations in both effectiveness and visibility.

If scientific institutions aim to improve the visibility of retractions, replications, and critical letters, several interventions could be considered. One approach is to enhance the prominence of these mechanisms on journal websites. For example, journals could make links to criticisms more visible on article landing pages or flag associated criticisms and commentaries in search engine results. Retractions, which are currently the most visible of these mechanisms, serve as a model due to their clear and prominent labeling.
Another approach involves implementing automated checks during the journal submission process. Such systems could flag references to retracted, replicated, or criticized studies, prompting authors to evaluate and justify these citations while providing referees with relevant information for review. Similar measures have already been proposed to increase awareness of retracted studies~\cite{BarIlan_2017_post-retraction, BornemannCimenti_2015_retracted}, but could be extended to linking critical letters, commentaries, and replications.
We argue that criticism is fundamental to scientific progress, and steps must be taken to ensure it is not overlooked. 
Enhanced visibility and thoughtful citation practices can help ensure that scholars engage with references alongside their full intellectual context.

%
%
\section*{Methods and Materials}

\subsection*{Data}
Data is sourced from the 12/2021 snapshot of the Microsoft Academic Graph (MAG)~\cite{sinha2015overview}.
MAG indexes basic paper metadata, disambiguated author profiles, and an ontology defining the hierarchy of disciplines, sub-disciplines, and topics, each of which multiple can be mapped to each paper. 
Each paper in MAG is tagged with a set of hierarchical \textit{field} categories based on its content and metadata; we consider only tags at the highest level of aggregation, representing 22 distinct fields.
The MAG citation network has low coverage for papers published in PRL, and so we supplement these with citation links taken from the 2020 version of the \textit{American Physical Society} citation dataset.
Further, we also source certain paper-level characteristics from SciSciNet~\cite{lin_2023_sciscinet}, a public data lake for science of science research. 

Critical letters from the journals \textit{Science}, \textit{Physical Review Letters} (\textit{PRL}), and \textit{Physical Review A-E} are identified by consistent naming conventions in the form of \textit{``Comment on ``Reduced Dynamics Need Not Be Completely Positive''}.
We query these directly from MAG to identify 480 critical letters and 404 targeted papers from \textit{Science} and 1,831 critical letters and 1,682 targeted papers from \textit{PRL}.
The journals \textit{Nature} and the \textit{Proceedings of the National Academy of Sciences} (\textit{PNAS}) do not use consistent naming conventions and cannot be identified from bibliometric metadata alone;
instead, we use their websites' search functionality to retrieve a list of documents that are of the equivalent type to a critical letter (``matters arising'' for \textit{Nature}, ``letters'' for \textit{PNAS}).
Using links embedded within each article's landing page we identify the DOI of the targeted paper. 
We consider only papers published between 2000 and 2020.
This results in 238 critical letters across 185 targeted publications for the journal \textit{Nature}, and  595 letters across 517 targeted publications for \textit{PNAS}. 
For all journals, papers that were eventually retracted are excluded. 

\subsection*{Paper characteristics}
Each paper in our dataset is characterized along a variety of metrics. 
The first of these is its impact; in all cases, we consider the t-year citation impact, defined as the total number of citations accumulated by a paper within a number of years, $N$;
unless otherwise stated, we use $t=3$, though generally the metric is highly correlated across values of $t$. 

Next, we compute a measure of the interdisciplinarity of papers;
following past bibliometric literature~\cite{chen_2015_interdisciplinary}, we operationalize this as the diversity of their references or of the papers that cite them.
Specifically we use the Simpson's Diversity Index~\cite{simpson_1949_diversity} which measures the balance and variety of categories represented in a population. 
The index is formalized as $\sum{(\frac{n}{N})^{2}}$ where $n$ is the number of members of a particular category, and $N$ is the total number of members across all categories. 
For each of our papers, the population is the set of its references (or citations), and the categories are the high-level MAG field categories assigned to these publications; in the case that a paper is assigned to multiple categories, it appears multiple times for each. 
Only papers with at least 10 references or 10 citations are considered.

A measure of \textit{novelty} is also computed for each paper. 
Here, novelty is defined based on a widely-used bibliometric measure that captures the atypicality of a paper's citations~\cite{uzzi_2013_atypical}. 
While \textit{SciSciNet} pre-computes this measure for records in MAG, we opt to implement the measure ourselves in order to include citation links from the \textit{American Physical Society} bibliographic dataset.
The measure is implemented by counting the total number of co-occurrences of journal pairs appearing in the citation lists of papers and comparing them against 10 null models in which citations are randomly distributed.
Z-scores are computed between the real and null values.
Then, each paper is represented as a set of Z-scores corresponding to each pair of journals that appear in its reference list.
The 10th percentile of a paper's Z-scores (that is, the 10\% quantile of their distribution of journal pair Z-scores) is used as its measure of novelty.
Only papers with at least 10 references are considered.

For the sake of consistent interpretation across metrics, we consider the reverse Simpson's index and the reverse novelty score, which larger values correspond to greater interdisciplinarity or greater novelty.

\subsection*{Matching}
For each paper targeted by a critical letter (treatment) we identify a matched set of peer papers with similar characteristics from the same journal (control). 
The first characteristic considered in a match is the paper's high-level field category. 

In the case where a paper is assigned to multiple fields, one is selected at random and a dummy variable set that notes it as multi-disciplinary. 
Papers must have an exact field match. 
Next we record the date of publication using the \textit{year}; the match tolerance for the year is a tunable parameter that we explore for papers in Table~\ref*{si:tab:paper-matched-ttest} and for authors in Tables~\ref*{si:tab:authors-matched-productivity-ttest} and ~\ref*{si:tab:authors-matched-impact-ttest}.
Because papers published earlier in the year have longer to accumulate citations when calculating impact, we also encode a variable for the \textit{quarter} which must exactly match. 
Finally, we match based on a paper's citation impact.
We implement two different matching policies for impact to be used in separate analyses. 
The first, used for the analysis presented in Fig.~\ref*{fig:paper-features}, simply matches based on the 3-year citation impact of papers.
The second, used for the analysis presented in Fig~\ref*{fig:matched-results-impact}, aims to match based on citations accumulated until the year of publication of the critical letter; for example, if a paper published in 2010 received a letter in 2014, then matching is performed based on 4-year impact.
If a critical letter was published in the same year as the original, then we use 1-year citations.
Papers must have at least 5 citations to be included in matching. 
Matching tolerance is measured as percentage point difference for log-normalized impact; the exact tolerance is a tunable parameter.

The authors of papers targeted by critical letters are also matched to a comparable control population. 
Matching is performed only for those we define as lead authors, specifically the first and last authors, who tend to have made the largest contributions to a paper and have the most ownership~\cite{lariviere_2016_division}, with first authors primarily executing a project whereas last authors more often serve as the conceptual lead~\cite{lariviere_2021_division}.
For each treatment paper, we collect all papers published in the same journal within a 4-year time window of the original from which we extract all their first and last authors.
A set of characteristics is computed for each candidate author based on these authors' publication histories.
Rather than match authors exactly based on time lag between the targeted paper and critical letter, we assume a standard lag of one year for all candidate authors for measuring changes in impact, which is the majority value among the treatment papers. 
We attempt to match authors based on field, age, productivity, and average impact of their papers. 
An author's field is defined as the modal high-level field appearing in papers across their career history. 
Career age is defined as the number of years between the author's first publication and the date of the critical letter (or equivalent time, for the candidate authors)
Productivity is operationalized as the average yearly fractionalized productivity in the five years preceding the letter; for example, a paper co-authored with 5 others will count for $\frac{1}{5}$ towards this count. 
Further, we calculate their lead productivity as the yearly average sum of papers in which they were a lead (first or last) author. 
We also consider the average field-normalized 3-year citation impact of papers published in the preceding 5 years, where normalization is performed based on the average impact of all papers in MAG with the same granular (level 1) field tag; in the case that a paper has multiple tags, the average is taken across each. 
The match tolerance for productivity and impact is tunable parameters defined as percentage point difference in the log-transformed value. 

\begin{figure}[p!]
    \centering
    \includegraphics[width=1.0\linewidth]{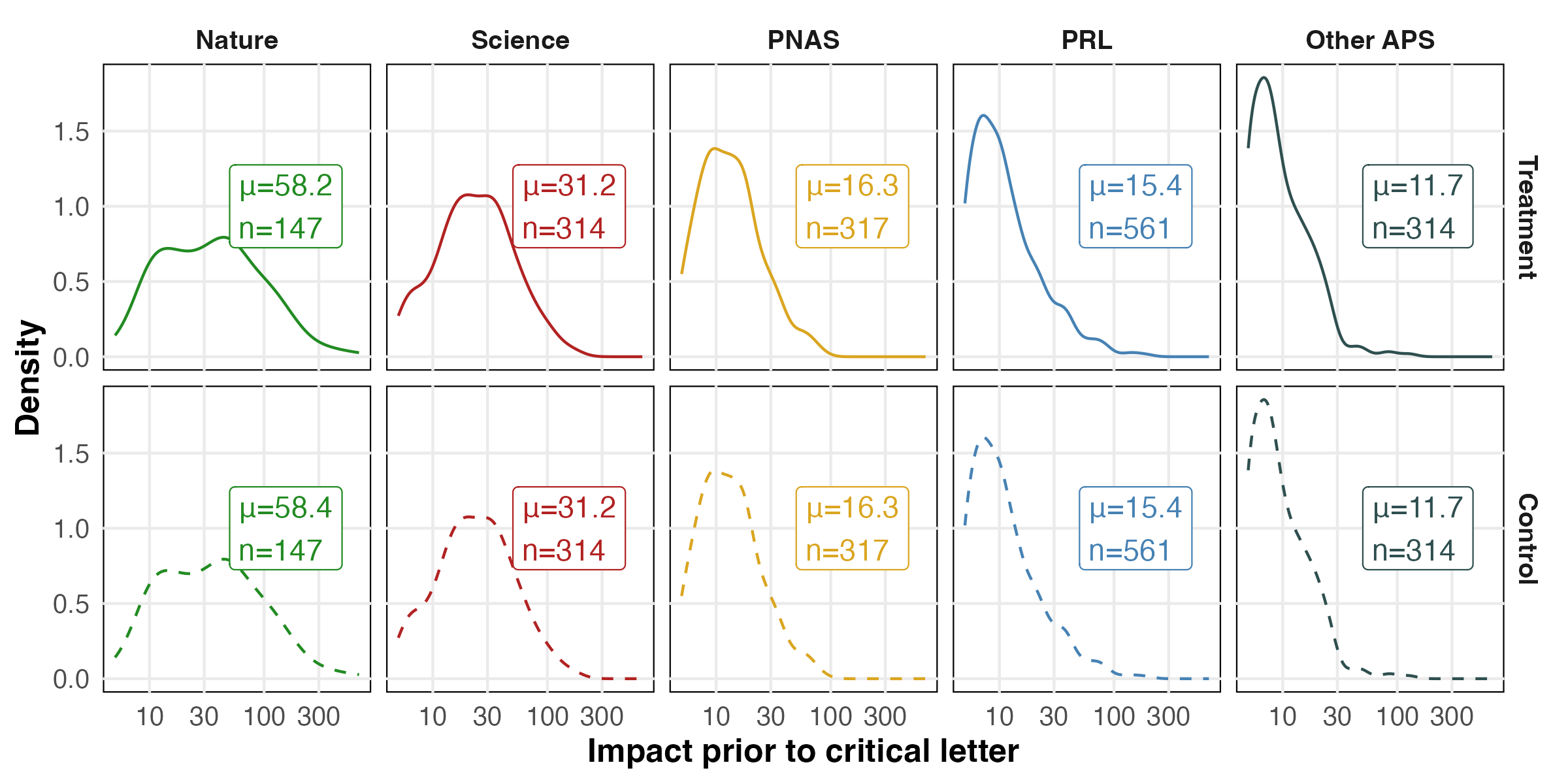}
    \caption{
    \textbf{Matches for papers by paper impact are high quality.}
    The distribution of citation impact of papers in the treatment group (top), and for the matched population of those with similar impact, field, and publication year (bottom). 
    Inset in each figure is the mean value of the distribution ($\mu$) and the number of matched observations ($n$). 
    The results show a high degree of similarity between the treatment population and their matched counterparts, suggesting a high quality. 
    }
    \label{si:fig:diagnostics}
\end{figure}

In all cases, matching is implemented in R's \textit{MatchIt} package~\cite{matchit_2011}. 
The technique used is propensity score matching, and for each record in the treatment, only the one nearest candidate is used.
The resulting matches are high quality, with a high degree of similarity between the matched and treatment populations (Fig.~\ref*{si:fig:diagnostics}, Tables~\ref*{si:tab:first-author-match-quality}-\ref*{si:tab:last-author-match-quality}).

Matching cannot fully account for the heterogeneity across the treatment populations.
To mitigate this issue, all analyses consider within-record comparisons. 
For papers, we compute the ratio of citations accumulated before the critical letter to those accumulated 3 years after; for example if a paper received 10 citations by the time it received a letter, and after three more years accumulated a total of 40 citations, then the ratio would be $\frac{40}{10} = 4.0$. 
We compute the same ratio for authors based on the ratio of their average field-normalized impact in the five years before and after the receipt of the critical letter.

\subsection*{Paper similarity}
To represent similarities between citing and cited papers, we use SPECTER~\cite{Singh2022SciRepEvalAM, cohan-etal-2020-specter}, a neural-network language model capable of generating general-purpose vector representations of scientific documents based on their text content. 
The model is initialized from the SciBERT language model~\cite{beltagy-etal-2019-scibert} but has an additional pre-training objective incorporating citation data to encode relatedness between citing and cited papers.
Thanks to this additional training that considers relationships between documents, SPECTER generally outperforms SciBERT, which is trained only on language modeling.

We use SPECTER to generate vector representations for papers that were targeted by critical letters for those which cite them. 
SPECTER is trained using both titles and abstracts, and generally expects both when generating embeddings.
Microsoft Academic Graph, however, does not index structured paper abstracts (instead, only containing bag-of-words representations), and even for other databases which provide abstracts (such as Dimensions) coverage can be low and biased. 
Given this, we generate embeddings using only titles, accepting any loss of performance under the assumption that general trends will remain unaffected. 

Similarity between the cited and citing papers is calculated based on the cosine similarity between their SPECTER vector representations.
Rather than report raw similarities, which can be difficult to interpret, our analysis instead relies on relative measures such as the percentile rank of similarity among the citations to a paper.

%
%
%
%
\subsection*{Acknowledgments}
We are grateful for support from the NSF (Grant \#2219575).
We thank the wonderful research community at the Center for Complex Network Research at Northeastern University.
For their helpful comments and feedback we thank Yian Yin, Brian Uzzi, Junming Huang, Alex Gates, Baruch Brazel, Filippo Radicchi, Jisung Yoon, Seokkyun Woo, and Zhouming Wu.

\subsection*{Author Contributions}
B.C, D.M, A.-L.B. designed research;
B.C., D.M., and Y.L performed research; 
B.C. and D.M analyzed data; 
and B.C, D.M. and A.-L.B.wrote the paper.

\subsection*{Declaration of competing interests}
A.-L.B. is co-scientific founder of and is supported by Scipher Medicine, Inc., which applies network medicine strategies to biomarker development and personalized drug selection, and founder of Naring, Inc. which applies data science to health. The remaining authors declare no competing interest.

\clearpage
\printbibliography

\clearpage

\begin{center}
    \LARGE Supporting Material: The origin, consequence, and visibility of criticism in science
    
    \vspace{0.5cm}
    \large
    Bingsheng Chen\textsuperscript{1,a}, Dakota Murray\textsuperscript{1,a}, Yixuan Liu\textsuperscript{a}, Albert-L\'{a}szl\'{o} Barab\'{a}si\textsuperscript{a,b,c}
    
    \vspace{0.5cm}
    \large
    \textsuperscript{a}Network Science Institute, Northeastern University, Boston, MA, 02115\\
    \textsuperscript{b}Department of Medicine, Brigham and Women’s Hospital, Harvard Medical School, Boston, MA, 02115\\
    \textsuperscript{c}Department of Network and Data Science, Central European University, Budapest 1051, Hungary\\
    \textsuperscript{1}Authors contributed equally to this work\\
    \vspace{0.5cm}
    \date{\today}
\end{center}

\maketitle 

\clearpage
\section*{Supplementary Information}
\beginsupplement
\begin{refsection}

%
%

\paragraph*{S1 Text}
\label{si:text:journal-selection}
\textbf{Journal selection:}
Critical letters were sourced on a journal-by-journal basis. 
Initially, we considered the list of journals analyzed in a previous study~\cite{radicchi_2012_comment}, which included \textit{Nature}, \textit{Science}, \textit{Physical Review Letters}, \textit{Physical Review A-E}, \textit{The New England Journal of Medicine}, \textit{Geology}, \textit{The Journal of Chemical Physics}, \textit{Water Resource Management}, and \textit{Environmental Science \& Technology}. 
However, after reviewing each journal, we narrowed our focus to a smaller subset divided into \textit{elite generalist} and \textit{specialist} journals.

The \textit{elite multi-disciplinary} journals in our study are \textit{Nature}, \textit{Science}, and \textit{Proceedings of the National Academy of Sciences} (\textit{PNAS}).
These journals are characterized by publishing papers in many fields, maintaining a broad and diverse readership, and being high-profile, representing a strong case for the visibility of critical letters.

Our \textit{specialist} journals include \textit{Physical Review Letters} and \textit{Physical Review A-E}. These journals, published by the American Physical Society, represent the broad physics community, reducing variability in editorial practices and disciplinary norms. 
To limit heterogeneity, we excluded journals like \textit{Geology}, \textit{Water Resource Management}, and \textit{The Journal of Chemical Physics}, where differences in editorial standards and disciplinary culture could confound analysis and interpretation.

We also excluded certain other specialist journals from our analysis.
For instance, we initially considered the \textit{New England Journal of Medicine} and the \textit{Journal of the American Medical Association}. 
However, a manual review revealed that only a minority of their published letters criticize original research, with most serving as commentaries on broader issues, applications, or field practices.
This is to say that these commentaries are not \textit{explicit} criticism: the mere presence of a commentary provides little information to the reader about its content.
Since this study focuses exclusively on criticism of original research, identifying relevant letters among thousands of publications in each of these journals was deemed infeasible within the scope of this work.
Even if they could be identified, the fact that criticisms are mixed with commentaries, that they tend to have a smaller word count, and that they tend not to provide additional evidence may make them incomparable to critical letters from other journals. 
This style of article commentary is commonplace among medical journals, and should be dealt with carefully in future analyses of criticism in science.

\paragraph*{S2 Text}
\label{si:text:exploratory-annotation}
\textbf{An exploratory study to characterize critical letters:}

We conducted an exploratory pilot study to characterize the nature of critical letters. 
The study aims to address two specific questions. 
First, to what extent do critical letters represent valid instances criticism? 
While such letters are broadly intended to engage critically with published work, the policies, cultures, and expectations for letters vary across journals. Our objective is to assess the degree to which these letters can be considered actual criticism. 
Second, are these critical letters substantiating their critiques with additional evidence? 
Differentiating between these is challenging, as all critical letters employ strong rhetoric and argumentation, which are substantive in themselves. 
However, many letters go further by supporting their critiques with novel evidence, such as replicating experiments, presenting new data, or conducting fresh analyses. 
We note that arguments grounded only in theory can be equally substantive.
However, we consider the pretense of new evidence as a reasonable proxy for substantive criticism.

To categorize critical letters as valid, we developed a rudimentary annotation scheme. Annotating critical letters is inherently challenging. 
They are often written in dense disciplinary jargon, and understanding the significance of specific statements requires substantial domain knowledge. 
Given these difficulties, our scheme is necessarily coarse-grained. 
We classify a letter as a valid instance of criticism if it presents arguments that raise uncertainty about at least one epistemic claim made in the original paper. 
Validity is coded as a binary variable (true/false).
Importantly, we do not differentiate between levels of criticism severity: a minor critique of a secondary claim and a rejection of a study’s primary findings are both treated as valid. 
Letters deemed invalid typically fall into one of several categories, such as those aiming to expand on the original study, offering commentary, or critiquing non-epistemic aspects, like open science practices. 
While we highlight examples of these letters in S3 Text, we do not systematically categorize these distinctions. 
Determining validity is often straightforward based on the abstract or opening paragraph, though in ambiguous cases, annotators rely on their best judgment.

A similar scheme was used to annotate whether a letter introduces new evidence.
Here, we define new evidence as any empirical work presented in the critical letter that has not been previously published.
This evidence should represent non-trivial effort on the side of the letters' authors, such as replication of portions of the target paper, collection of new data, analysis of third party data, and so on. 
Excluded from this definition are use of reference images available to all researchers (e.g., relief maps or other geographic imagery, images of biological specimens available in scientific archives), or which are taken from previous studies.
Like validity, this categorization is coarse-grained, and we do not distinguish between types of evidence, their quality, or their relevance. 
Identifying new evidence is generally straightforward, with many letters containing clear cues. 
For instance, figures or tables and their captions may signal the presence of new evidence, though they might also serve as references.
Textual indicators, such as phrases like "we reanalyze the data..." or mentions of data and methodologies in the main text, figure captions, article notes, data availability statements, or supplementary materials, also provide clues. 
In cases where these cues are absent, annotators rely on their best judgment when it is evident that new evidence is introduced.

One of the authors of this paper annotated a subset of critical letters from three journals: \textit{Nature}, \textit{Science}, and \textit{PNAS}. 
\textit{APS} papers are not annotated as these letters are often terse and difficult for parse for the non-physicist annotator, whereas papers in these more generalist journals tended to be more intelligible.
For \textit{Nature} and \textit{Science}, we searched their respective websites for all critical letters ("Matters Arising" in \textit{Nature} and "Technical Comments" in \textit{Science}) published during 2019, 2020, and 2021, resulting in 50 letters from \textit{Nature} and 62 from \textit{Science}. 
\textit{PNAS}, which publishes substantially more critical letters (called "Letters") annually, required a smaller time window to obtain a reasonable sample; we therefore annotated 102 letters from 2019. In total, a single annotator reviewed and coded all 214 letters from the three journals.

The results of the annotation are included with the research data published alongside this manuscript.
Here, we highlight some of noteworthy critical letters and the labels that were assigned,

\begin{itemize}
    \item Soltesz et al., (\textit{Nature}, 2020)~\cite{soltesz_2020_interventions} is a \textbf{valid criticism} that questions the the estimates of effectiveness for certain non-pharmaceutical interventions of COVID-19. 
    It is also \textbf{introduces new evidence}.
    The authors of the critical letter replicate the original analysis with new model parameters, demonstrating that that the results are sensitive to such choices.

    \item Haibe-Kains et al., (\textit{Nature}, 2020)~\cite{HaibeKains_2020_transparency} presents criticism of a study that evaluates the AI systems towards screening of breast cancer. 
    However, the critical letter does not critique the content of the original study, but rather its transparency. 
    Namely, the targeted study does not provide the data, parameter selection, and code necessary to reproduce its findings.
    We consider this to be an instance of \textbf{invalid} criticism, as it does not critique the truth claims of the original study. 
    Further, while this article provides tables to help illustrate their argument, we consider it to \textbf{not introduce new evidence}, as they do not represent new analysis, data, or significant effort on the part of the letters' authors.
    
    \item Rice, Craig, \& Dyda (\textit{Science}, 2020)~\cite{rice_2020_rna} outlines critique a methodology paper that introduces a technique for research in the bio-sciences. 
    This is a unique circumstance, as the original paper introduces a method rather than make a truth claim. 
    However, the critical letter notes necessary nuances and issues that make the method less widely applicable than is presented in the original paper.
    For this reason, we consider this to be an instance of \textbf{valid criticism}.
    The letter, however, \textbf{does not introduce new evidence}, relying instead on citations to previous literature to make their argument.

    \item Heuvel \& Tauris (\textit{Science}, 2020)~\cite{vandenHeuvel_2020_black-hole} targets a research article claiming evidence that a particular celestial body is a black hole. 
    The critical letter does not reject this claim outright. 
    Rather, they argue that the target papers' data are consistent with multiple possible conclusions.
    This is highlighted by their closing statement ``We conclude that the unseen companion of 2MASS J05215658 might not be a black hole.''. 
    We still consider this a \textbf{valid instance of criticism}, having implications for the significance, interpretation, and certainty of the original study.
    It \textbf{does not introduce new evidence}.

    \item Rosen (\textit{PNAS}, 2019)~\cite{rosen_2019_temperature} is \textbf{valid criticism}, as evidenced by an atypically bold opening statement: ``I believe that all of the numerical results cited in this article are wrong, because the methodology is not valid.''
    However, this letter is considered to \textbf{not introduce new evidence}.
    The argument of this letter rests on noting flaws in the target paper's methodology that the letter writer claims results in incorrect findings.

    \item Loisel et al., (\textit{PNAS}, 2019)~\cite{loisel_2019_soils} is an \textbf{not a valid criticism} as it is responding to an opinion piece, as noted in the opening sentence. 
    We note that such a letter is not present in our primary analysis, because we limit to only critical letters that target research articles. 
\end{itemize}

The annotation results are summarized in Table~\ref*{si:table:annotation}.
The majority of critical letters across all three journals were annotated as valid, the highest of 95\% in \textit{Science}, followed by 90\% in \textit{Nature}, and the lowest of 75\% for \textit{PNAS}. 
\textit{PNAS} publishes a number of commentaries.
A smaller portion of these critical letters introduced new evidence, 68\% for both \textit{Nature} and \textit{Science}, and a minority of 38\% for \textit{PNAS}.
Many of the letters in \textit{PNAS} rely on argumentation and citation to previous literature to substantiate their critique; these are still valid criticism, but by the definition of our annotation scheme do not meet the threshold of introducing new evidence. 

We re-iterate that this annotation study is exploratory in nature, designed to provide preliminary systematic evidence supporting a few narrow claims about the characteristics and substance of critical letters. 
We recognize the limitations of our coarse-grained annotation scheme and the relatively small sample size.
Future research should build on this approach by developing more nuanced and comprehensive annotation frameworks that better capture the diversity of critical letters.
Additionally, future studies should involve multiple annotators and assess inter-rater reliability to enhance robustness.
Despite its limitations, this exploratory study represents a significant manual effort and offers valuable context for interpreting the findings presented in this manuscript.

\paragraph{S3 Text}
\label{si:text:marginal-prob}
\textbf{Marginal probabilities of author demographics on likelihood of criticism:}
We examine the likelihood that author demographic characteristics, in particular their \textit{gender}, \textit{seniority}, and \textit{university prestige}, have on the likelihood of the receipt of a critical letter. 
Gender is inferred based on the given name of researchers in MAG following previous work~\cite{huang_2020_gender};
as a note of caution, we note the inherent limitations of name-based gender inference, including the inability to handle non-binary gender identities, challenges at distinguishing gender from anglicized versions of names common in East-Asian countries and cultures, and spotty data on given name in bibliometric metadata, leading to many missing values that limit the scale of analysis. 
Due to these limitations, we consider our analysis of gender exploratory, pending further analysis of self-reported data. 

Seniority is implemented as a dummy variable indicating that a researcher's career age---the number of years between their first publication and the publication of the target paper---is at least ten years. 

The prestige of a researchers' university is determined using the Leiden rankings~\cite{van_eck_2023_leiden}, a systematic ranking of global universities based on bibliometric indicators such as publication count, impact, and more.
We limit to universities which have published at least 1000 papers a year between 2000 and 2020, and rank them by the proportion of their papers which are in the top 5\% of all papers in terms of field-normalized impact.
Universities which were ranked within the top 30 within any year of the time period are considered elite, comprising 53 total institutions.
We note that the Leiden Rankings are an imperfect proxy for prestige---some institutions which might not be considered prestigious, particularly hospitals, are over-represented due to a high proportion of high-impact publications, whereas others are absent;
Still, the Leiden Rankings offers an objective measure of ranking global institutions. 
The 53 universities in this top list are manually matched to a GRID ID which allows integration with the Microsoft Academic Graph. 

\begin{figure}
    \centering
    \includegraphics[width=1.0\linewidth]{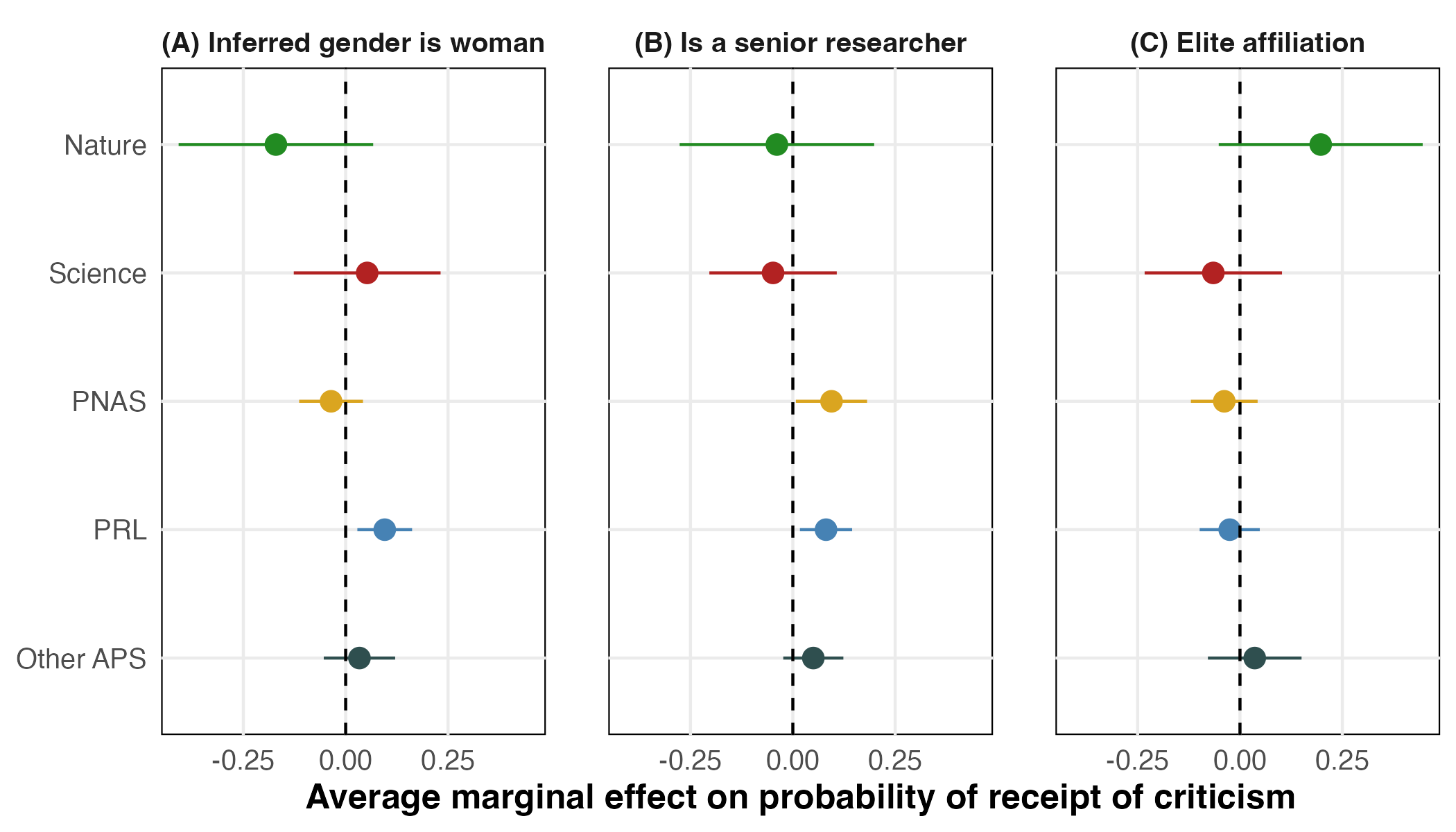}
    \caption{
    \textbf{Average marginal probabilities of (A) first author gender, (B) first author seniority, and (C) first author affiliation prestige, on likelihood of receipt of criticism.}
    Shown here for first authors only. 
    The data used in this graph comprises papers targeted by critical letters along with matched peer papers, which underlies Fig~\ref*{fig:paper-features}. 
    Marginal probabilities are calculated for a logistic regression model with the dependent variable representing whether a paper received a critical letter. 
    Control variables include the log-scaled field-normalized impact, the number of authors on the paper, and the year of publication. 
    Each point corresponds to the average marginal effect of each independent variable, including whether the first author is a woman (inferred by first name), whether they are a senior researcher (a career age greater than 10 years), or are affiliated with an elite university (measured by the Leiden rankings). 
    A positive average marginal effect represents that the variable is associated with an increase in likelihood of belonging to the set of papers that received criticism. 
    }
    \label{si:fig:margins_first-auhor}
\end{figure}

\begin{figure}
    \centering
    \includegraphics[width=1.0\linewidth]{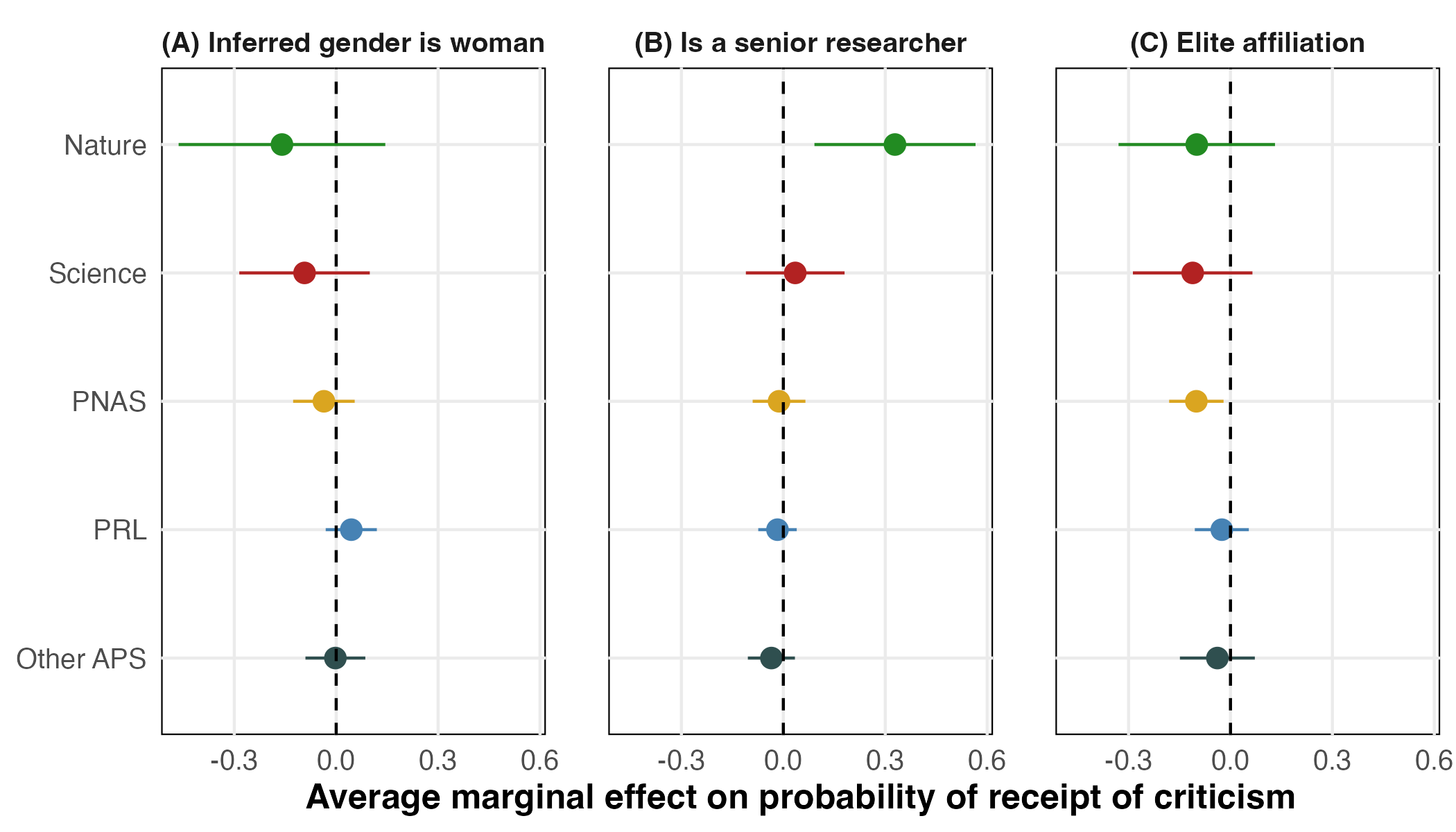}
    \caption{
    \textbf{Average marginal probabilities of (A) last author gender, (B) last author seniority, and (C) last author affiliation prestige, on likelihood of receipt of criticism.}
    Shown here for last authors only. 
    The data used in this graph comprises papers targeted by critical letters along with matched peer papers, which underlies Fig~\ref*{fig:paper-features}. 
    Marginal probabilities are calculated for a logistic regression model with the dependent variable representing whether a paper received a critical letter. 
    Control variables include the log-scaled field-normalized impact, the number of authors on the paper, and the year of publication. 
    Each point corresponds to the average marginal effect of each independent variable, including whether the first author is a woman (inferred by first name), whether they are a senior researcher (a career age greater than 10 years), or are affiliated with an elite university (measured by the Leiden rankings). 
    A positive average marginal effect represents that the variable is associated with an increase in likelihood of belonging to the set of papers that received criticism. 
    }
    \label{si:fig:margins_last-auhor}
\end{figure}

Using these data, we set out to test the question of whether gender, seniority, or prestige are associated with an increased likelihood of receiving of a critical letter. 
To test this question we use the dataset underlying Fig~\ref*{fig:paper-features}, consisting of pairs of papers that received a critical letter matched to others of the same journal which did not.
Papers which have more than 20 authors are excluded.
For each journal, and for each of first and last authors, we fit a logistic regression model consisting of a binary variable indicating receipt of a critical letter as the dependent variable, and the gender, seniority, and prestige dummy variables as independent variables. 
We also include control variables such the field-normalized and log-scaled 3-year citation impact of the paper, a discretized count of the number of authors (one of ``1'', ``2-5'', ``5-10'', or ``10-20''), and the year of publication. 
Once fit, we compute the average marginal effect of each of the three independent variables in the model, with results shown for first authors in Fig~\ref*{si:fig:margins_first-auhor}, and for last authors in Fig~\ref*{si:fig:margins_last-auhor}. 
Values can be interpreted as percentage increase or decrease that being female, being a senior researcher, or being affiliated with an elite university has on the likelihood of belonging to the subset of papers which received a critical letter, compared to the control group of matched papers. 

The results illustrate that author demographics have little notable effect on their receipt of a critical letter. 
Few results have confidence intervals which do not cross zero;
here we focus on those that do or come close. 
In terms of gender, we observe a slight trend such that papers in which the last author is inferred as a woman are 6.8 percentage points more likely than men to belong to the criticized group, which we also observe for first authors (an increase in likelihood of 7.4).
For seniority, we observe that having a career age greater than ten years is associated with a slight increase in the likelihood of receiving a criticism for first authors at \textit{PNAS} (10.4 points) and \textit{PRL} (7.4 points), yet this is not true for last authors, for whom the strongest effect is for \textit{Nature} (34 points);
we note however that \textit{Nature} had the fewest number of observations in the final model, and so this result may be suspect. 
To our surprise---as we had assumed that faculty from elite universities would be less likely to receive a critical letter---we observe no notable effects for elite university affiliation for either first or last authors. 
These findings suggest that, in certain circumstances, gender and seniority are associated with likelihood of receipt of a critical letter, yet more investigation is necessary to provide confirmatory support and to understand differences between journals. 

\paragraph{S4 Text}
\textbf{Disruptiveness:}
We specifically avoid the use the the Disruptiveness indicator in this work. 
It may be interesting to know whether papers that are targeted by criticism tend to be more disruptive than other papers, and such calculations are readily available in SciSciNet.
However, the measure has itself been been the subject of criticism on both methodological and theoretical grounds~\cite{funk_2017_disruption,leibel_2024_disruption}. 
As such, we avoid analysis of it here, but encourage further work to make use of this (or similar) indicator when their validity and significance has been assured. 

\clearpage

\clearpage

\clearpage

\clearpage

\clearpage

\clearpage

\clearpage

\clearpage

\clearpage

%
%
%

%
%
%
%
\clearpage

\begin{table}[p]
\centering
\caption{Counts of critical letters and articles meeting inclusion criteria across all selected journals.}
\begin{tabular}{lrr}
  \toprule
Journal & Articles & Critical letters \\ 
  \midrule
Nature & 26138 & 179 \\ 
  Science & 24010 & 396 \\ 
  PNAS & 69515 & 513 \\ 
  PRL & 57422 & 834 \\ 
  PR-A & 43234 & 205 \\ 
  PR-B & 101651 & 221 \\ 
  PR-C & 18542 & 37 \\ 
  PR-D & 53549 & 98 \\ 
  PR-E & 40636 & 185 \\ 
   \bottomrule
\end{tabular}
\end{table}

\clearpage
\begin{table}[p]
\centering
\caption{
  Nagelkerke $R^2$ for a logistic regression model fit of all papers in each journal. 
  The dependent variable is a binary indicating whether the paper was targeted by a critical letter.
  The independent variables include the Year of publication along with the log-transformed 2-year impact. 
  We consider both the raw and field-normalized impact.
  The differences between the two measures are mostly negligible, with the largest different observed for \textit{Science}. 
} 
\label{tab:psuedor2}
\begin{tabular}{lrr}
  \toprule
Venue & 2-Year Impact & Normalized 2-Year Impact \\ 
  \midrule
Nature & 0.078 & 0.083 \\ 
  Science & 0.044 & 0.061 \\ 
  PNAS & 0.183 & 0.180 \\ 
  PRL & 0.030 & 0.032 \\ 
  Other APS & 0.013 & 0.015 \\ 
   \bottomrule
\end{tabular}
\end{table}

\clearpage
\begin{table}[p] \centering 
  \caption{
    Results of manual annotation of critical letters as valid instances of criticism and as whether they introduce new evidence. 
    The table includes the total count of documents annotated across the journals \textit{Nature}, \textit{Science}, and \textit{PNAS}, along with the percentage of each annotation category. 
    Validity is determined based on whether the letter includes statements that challenge the certainty of claims made by the targeted paper.
    Introduction of new evidence is determined based on whether the letter includes novel work not previously published such as re-analyses or replications of the original work or analyses of other data. 
  } 
  \label{si:table:annotation} 
\begin{tabular}{@{\extracolsep{5pt}} lrrr} 
\\[-1.8ex]\hline 
\hline \\[-1.8ex] 
Journal & Count & \% Valid & \% New Evidence \\ 
\hline \\[-1.8ex] 
Nature  & 50 & 0.90 & 0.68 \\ 
Science & 62 & 0.95 & 0.68 \\ 
PNAS & 102 & 0.73 & 0.38 \\ 
\hline \\[-1.8ex] 
\end{tabular} 
\end{table}

\clearpage
\begin{table}[p]
\centering
\caption{
    Table representation of data from Fig.~\ref*{fig:paper-features} with results of 1-way Komoragov-Smirnof statistical test. 
    Shown for each paper-level metric and journal. 
    Mean Rank (referred to in Fig.~\ref*{fig:paper-features} as $mu$ and in-text as $mu_{rank}$) denotes the average percentile rank of papers targeted by criticism.
    The test statistic and p-value are the results of a 1-way Kolmogorov–Smirnov test that compares the distribution of percentile ranks for targeted papers against a uniform distribution. 
    A low p-value indicates that targeted papers are concentrated among higher ranks, rather than being randomly distributed across the journal.
}
\label{si:table:paper-features-1sKS} 
\begin{tabular}{llrrrr}
  \toprule
Metric & Venue & N & Mean Rank & Test Statistic & P Value \\ 
  \midrule
Impact & Nature &  179 & 67.0 & 0.2810 & 0.0000 \\ 
  Impact & Science &  396 & 66.2 & 0.2630 & 0.0000 \\ 
  Impact & PNAS &  513 & 67.7 & 0.2590 & 0.0000 \\ 
  Impact & PRL &  834 & 54.2 & 0.0780 & 0.0000 \\ 
  Impact & Other APS &  746 & 58.7 & 0.1360 & 0.0000 \\ 
  Ref. Diversity & Nature &  178 & 54.1 & 0.1010 & 0.0260 \\ 
  Ref. Diversity & Science &  376 & 54.2 & 0.0830 & 0.0060 \\ 
  Ref. Diversity & PNAS &  510 & 56.4 & 0.1210 & 0.0000 \\ 
  Ref. Diversity & PRL &  655 & 51.4 & 0.0540 & 0.0220 \\ 
  Ref. Diversity & Other APS &  618 & 50.3 & 0.0310 & 0.3080 \\ 
  Cite. Diversity & Nature &  179 & 54.2 & 0.1000 & 0.0290 \\ 
  Cite. Diversity & Science &  394 & 50.8 & 0.0520 & 0.1220 \\ 
  Cite. Diversity & PNAS &  503 & 57.6 & 0.1190 & 0.0000 \\ 
  Cite. Diversity & PRL &  727 & 51.2 & 0.0660 & 0.0020 \\ 
  Cite. Diversity & Other APS &  490 & 48.3 & 0.0170 & 0.7480 \\ 
  Novelty & Nature &  179 & 55.1 & 0.1200 & 0.0060 \\ 
  Novelty & Science &  394 & 56.5 & 0.1540 & 0.0000 \\ 
  Novelty & PNAS &  512 & 58.3 & 0.1540 & 0.0000 \\ 
  Novelty & PRL &  286 & 55.2 & 0.1100 & 0.0010 \\ 
  Novelty & Other APS &  330 & 55.2 & 0.0940 & 0.0030 \\ 
   \bottomrule
\end{tabular}
\end{table}

\clearpage
\begin{table}[p]
\scriptsize
\centering
\caption{
Table representation of data from Fig.~\ref*{fig:paper-features} with results of 2-way Komoragov-Smirnof statistical test. 
Shown for each paper-level metric and journal. 
Mean Rank (referred to in Fig.~\ref*{fig:paper-features} as $mu$ and in-text as $mu_{rank}$) denotes the average percentile rank of papers targeted by criticism.
The test statistic and p-value are the results of a two-sample, one-sided KS test compares the distribution of percentile ranks for targeted papers against the matched population of comparable papers. 
Matching is performed as described in \methods.
A low p-value suggests that targeted papers are concentrated among higher ranks compared to their matched counterparts.
}
\label{si:table:paper-features-2sKS} 
\begin{tabular}{llrrrrr}
  \toprule
Metric & Venue & N & Mean Rank (Criticism) & Mean Rank (¬Criticism) & Test Statistic & P Value \\ 
  \midrule
  Ref. Diversity & Nature &  178 & 51.3 & 44.3 & 0.2619 & 0.0560 \\ 
  Ref. Diversity & Science &  376 & 52.0 & 42.4 & 0.1798 & 0.0560 \\ 
  Ref. Diversity & PNAS &  510 & 54.8 & 46.2 & 0.1530 & 0.0000 \\ 
  Ref. Diversity & PRL &  655 & 51.1 & 44.6 & 0.1525 & 0.0000 \\ 
  Ref. Diversity & Other APS &  618 & 50.2 & 45.9 & 0.1091 & 0.0200 \\ 
  Cite. Diversity & Nature &  179 & 53.0 & 46.2 & 0.1860 & 0.2280 \\ 
  Cite. Diversity & Science &  394 & 52.0 & 42.9 & 0.2212 & 0.0060 \\ 
  Cite. Diversity & PNAS &  503 & 56.1 & 44.9 & 0.1805 & 0.0000 \\ 
  Cite. Diversity & PRL &  727 & 49.2 & 45.3 & 0.0838 & 0.0190 \\ 
  Cite. Diversity & Other APS &  490 & 49.1 & 46.2 & 0.1030 & 0.0300 \\ 
  Novelty & Nature &  179 & 55.5 & 46.3 & 0.2093 & 0.1530 \\ 
  Novelty & Science &  394 & 56.8 & 51.5 & 0.1553 & 0.0830 \\ 
  Novelty & PNAS &  512 & 58.1 & 53.8 & 0.0843 & 0.0800 \\ 
  Novelty & PRL &  286 & 59.4 & 49.2 & 0.2796 & 0.0010 \\ 
  Novelty & Other APS &  330 & 52.6 & 47.8 & 0.1150 & 0.2240 \\ 
   \bottomrule
\end{tabular}
\end{table}

\clearpage
\begin{table}[p]
\scriptsize
\centering
\caption{
    Table representation of data from Fig.~\ref*{fig:embedding} with results of 1-way Komoragov-Smirnof statistical test.
    Shown for each journal. 
    Mean Rank (referred to in Fig.~\ref*{fig:paper-features} as $mu$ and in-text as $mu_{rank}$) denotes the average percentile rank of papers targeted by criticism.
    The test statistic and p-value are the results of a 1-way Kolmogorov–Smirnov test that compares the distribution of percentile ranks for targeted papers against a uniform distribution. 
    A low p-value indicates that papers which co-cite the critical letter are among the most similar that cite the critized paper. 
}
\label{si:table:embedding-tests} 
\begin{tabular}{lrrrr}
  \toprule
Venue & N & Mean Rank & Test Statistic & P Value \\ 
  \midrule
Nature & 5106 & 59.3 & 0.1308 & 0.0000 \\ 
  Science & 7878 & 58.9 & 0.1296 & 0.0000 \\ 
  PNAS & 2458 & 57.5 & 0.1164 & 0.0000 \\ 
  PRL & 9901 & 52.5 & 0.0416 & 0.0000 \\ 
  Other APS & 4389 & 53.3 & 0.0518 & 0.0000 \\ 
   \bottomrule
\end{tabular}
\end{table}

%
%
%
\clearpage
\begin{table}[p]
\small
\centering
\caption{
\textbf{Counts of matched papers.} 
``Delay'' denotes the number of years after which cumulative impact is measured; higher values reduce the data's temporal scope and the number of candidates. 
``Impact $\pm \epsilon$'' specifies the tolerance for matching citation impact, calculated as the log-transformed fraction of cumulative citations normalized by the journal's average, with all values field-normalized. 
Matched papers must fall within the specified impact tolerance after a period corresponding to the lag between the treatment paper and its critical letter (e.g., a 2-year lag uses the 2-year impact). 
``Year $\pm \epsilon$'' indicates the maximum allowed difference in publication years between treatment and candidate papers. 
For each journal, each cell reports the raw number of matches and the percentage of total candidates meeting the criteria: sufficient data for impact measurement, treatment paper impact exceeding five citations, a lag under six years, and complete metadata. 
}
\label{si:tab:paper-matched-counts}
\begin{tabular}{llllllll}
  \toprule
Delay & Impact $\pm$ $\epsilon$ & Year $\pm$ $\epsilon$ & Nature & Science & PNAS & PRL & Other APS \\ 
  \midrule
2 & 5\% & 2 & 148 (84.6\%) & 339 (87.1\%) & 369 (80.2\%) & 572 (87.5\%) & 320 (74.6\%) \\ 
  3 & 5\% & 2 & 147 (84\%) & 314 (84\%) & 317 (83.4\%) & 561 (87.1\%) & 314 (77.1\%) \\ 
  4 & 5\% & 2 & 137 (79.2\%) & 303 (85.1\%) & 260 (81.2\%) & 552 (87.1\%) & 301 (77.8\%) \\ 
   &  &  &  &  &  &  &  \\ 
  3 & 5\% & 2 & 147 (84\%) & 314 (84\%) & 317 (83.4\%) & 561 (87.1\%) & 314 (77.1\%) \\ 
  3 & 10\% & 2 & 151 (86.3\%) & 335 (89.6\%) & 328 (86.3\%) & 570 (88.5\%) & 315 (77.4\%) \\ 
  3 & 15\% & 2 & 158 (90.3\%) & 343 (91.7\%) & 334 (87.9\%) & 572 (88.8\%) & 314 (77.1\%) \\ 
   &  &  &  &  &  &  &  \\ 
  3 & 5\% & 1 & 143 (81.7\%) & 291 (77.8\%) & 284 (74.7\%) & 558 (86.6\%) & 307 (75.4\%) \\ 
  3 & 5\% & 2 & 147 (84\%) & 314 (84\%) & 317 (83.4\%) & 561 (87.1\%) & 314 (77.1\%) \\ 
   \bottomrule
\end{tabular}
\end{table}

\clearpage
\begin{table}[p]
\centering
\caption{
\textbf{High quality of matches among first authors.}
The pooled average of several features for treatment and matched authors across each journal. 
Matches shown here are identified with a impact tolerance and productivity tolerance both set to 10\%.
Career age is the number of years between an authors' first publication and the period of the critical letter. 
Impact is calculated for each author as the average field-normalized 3-year citation impact of their papers published within the five years prior to the receipt of the critical letter.
Productivity is calculated for each author as their average yearly fractionalized number of publications, normalized by the average of all authors in the same venue.
Results show a high degree of similar in metrics between the populations.
}
\label{si:tab:first-author-match-quality}
\begin{tabular}{llrrr}
  \toprule
Journal & Type & Career Age & Impact & Productivity \\ 
  \midrule
Nature & Treatment & 14.929 & 2.575 & 0.726 \\ 
  Nature & Control & 14.486 & 2.566 & 0.728 \\ 
  Science & Treatment & 15.358 & 2.588 & 0.801 \\ 
  Science & Control & 15.044 & 2.585 & 0.799 \\ 
  PNAS & Treatment & 16.303 & 2.305 & 0.914 \\ 
  PNAS & Control & 15.794 & 2.320 & 0.918 \\ 
  PRL & Treatment & 14.514 & 1.386 & 0.814 \\ 
  PRL & Control & 13.964 & 1.388 & 0.814 \\ 
  Other APS & Treatment & 17.069 & 1.243 & 0.784 \\ 
  Other APS & Control & 16.836 & 1.247 & 0.785 \\ 
   \bottomrule
\end{tabular}
\end{table}

\clearpage
\begin{table}[p]
\centering
\caption{
\textbf{High quality of matches among last authors.}
The pooled average of several features for treatment and matched authors across each journal. 
Matches shown here are identified with a impact tolerance and productivity tolerance both set to 10\%.
Career age is the number of years between an authors' first publication and the period of the critical letter. 
Impact is calculated for each author as the average field-normalized 3-year citation impact of their papers published within the five years prior to the receipt of the critical letter.
Productivity is calculated for each author as their average yearly fractionalized number of publications, normalized by the average of all authors in the same venue.
Results show a high degree of similar in metrics between the populations.
}
\label{si:tab:last-author-match-quality}
\begin{tabular}{llrrr}
  \toprule
Journal & Type & Career Age & Impact & Productivity \\ 
  \midrule
Nature & Treatment & 22.871 & 2.714 & 0.851 \\ 
  Nature & Control & 23.137 & 2.706 & 0.854 \\ 
  Science & Treatment & 20.944 & 2.541 & 0.864 \\ 
  Science & Control & 20.996 & 2.539 & 0.865 \\ 
  PNAS & Treatment & 23.342 & 2.429 & 0.880 \\ 
  PNAS & Control & 23.584 & 2.425 & 0.878 \\ 
  PRL & Treatment & 20.527 & 1.466 & 0.742 \\ 
  PRL & Control & 20.384 & 1.466 & 0.742 \\ 
  Other APS & Treatment & 20.631 & 1.322 & 0.664 \\ 
  Other APS & Control & 19.748 & 1.324 & 0.665 \\ 
   \bottomrule
\end{tabular}
\end{table}

\clearpage
\begin{table}[p]
\scriptsize
\centering
\caption{
\textbf{T-Test comparing post-criticism impact growth of treatment papers against matched population.}
``Delay'' denotes the number of years after which cumulative impact is measured; higher values reduce the data's temporal scope and the number of candidates. 
``Impact $\pm \epsilon$'' specifies the tolerance for matching citation impact, calculated as the log-transformed fraction of cumulative citations normalized by the journal's average, with all values field-normalized. 
Matched papers must fall within the specified impact tolerance after a period corresponding to the lag between the treatment paper and its critical letter (e.g., a 2-year lag uses the 2-year impact). 
``Year $\pm \epsilon$'' indicates the maximum allowed difference in publication years between treatment and candidate papers. 
For each journal, the first value indicates the test statistic of a paired T-Test comparing the paper impact of the treatment and control.
Here, paper impact is defined as the ratio between the citations received by a paper after receipt of a critical letter compared
to those received prior; in the case of the control population we use an equivalent time lag.
The parenthetical next to the test statistic lists the p-value obtained from the test. 
}
\label{si:tab:paper-matched-ttest}
\begin{tabular}{llllllll}
  \toprule
Delay & Impact $\pm$ $\epsilon$ & Year $\pm$ $\epsilon$ & Nature & Science & PNAS & PRL & Other APS \\ 
  \midrule
2 & 5\% & 2 & 0.14 (p=0.093) & 0.03 (p=0.699) & 0.08 (p=0.208) & 0.08 (p=0.090) & -0.13 (p=0.036) \\ 
  3 & 5\% & 2 & 0.23 (p=0.132) & 0.01 (p=0.954) & 0.14 (p=0.217) & 0.07 (p=0.363) & -0.22 (p=0.011) \\ 
  4 & 5\% & 2 & 0.29 (p=0.221) & 0.18 (p=0.336) & 0.01 (p=0.930) & 0.07 (p=0.552) & -0.24 (p=0.051) \\ 
   &  &  &  &  &  &  &  \\ 
  3 & 5\% & 2 & 0.23 (p=0.132) & 0.01 (p=0.954) & 0.14 (p=0.217) & 0.07 (p=0.363) & -0.22 (p=0.011) \\ 
  3 & 10\% & 2 & 0.21 (p=0.150) & 0.16 (p=0.189) & 0.15 (p=0.146) & 0.08 (p=0.291) & -0.18 (p=0.035) \\ 
  3 & 15\% & 2 & 0.45 (p=0.001) & 0.18 (p=0.152) & 0.09 (p=0.411) & 0.05 (p=0.500) & -0.13 (p=0.126) \\ 
   &  &  &  &  &  &  &  \\ 
  3 & 5\% & 1 & 0.17 (p=0.253) & -0.01 (p=0.960) & 0.13 (p=0.284) & 0.05 (p=0.503) & -0.2 (p=0.020) \\ 
  3 & 5\% & 2 & 0.23 (p=0.132) & 0.01 (p=0.954) & 0.14 (p=0.217) & 0.07 (p=0.363) & -0.22 (p=0.011) \\ 
   \bottomrule
\end{tabular}
\end{table}

\clearpage
\begin{table}[p]
\scriptsize
\centering
\caption{
\textbf{Counts of matched authors.} 
``Authorship'' denotes the position of the authorship, either the first author or the last author. 
``Impact $\pm \epsilon$'' specifies the percentage tolerance for matching authors' citation impact, calculated as the log of their average 3-year field normalized impact of all papers published in the five years preceding receipt of the critical letter, normalized by the average across all authors in the same journal. 
``Productivity $\pm \epsilon$'' specifies the percentage tolerance for matching authors' productivity, calculated as their publication count, fractionalized by total authorship, normalized by the average across all authors in the same journal.
For each journal, each cell reports the raw number of matched records and, in the parentheses, the count as a percentage of all authors in the journal considered for matching
}
\label{si:tab:authors-matched-counts}
\begin{tabular}{llllllll}
  \toprule
Authorship & Impact $\pm$ $\epsilon$ & Productivity $\pm$ $\epsilon$ & Nature & Science & PNAS & PRL & Other APS \\ 
  \midrule
first & 5\% & 10\% & 34 (19.1\%) & 95 (24.2\%) & 80 (15.7\%) & 180 (21.6\%) & 48 (6.4\%) \\ 
  first & 10\% & 10\% & 41 (23\%) & 104 (26.5\%) & 82 (16.1\%) & 202 (24.2\%) & 76 (10.2\%) \\ 
  first & 15\% & 10\% & 47 (26.4\%) & 111 (28.2\%) & 87 (17.1\%) & 212 (25.5\%) & 85 (11.4\%) \\ 
  last & 5\% & 10\% & 66 (37.1\%) & 144 (36.6\%) & 171 (33.5\%) & 270 (32.4\%) & 106 (14.2\%) \\ 
  last & 10\% & 10\% & 77 (43.3\%) & 171 (43.5\%) & 181 (35.5\%) & 290 (34.8\%) & 145 (19.5\%) \\ 
  last & 15\% & 10\% & 83 (46.6\%) & 174 (44.3\%) & 183 (35.9\%) & 296 (35.5\%) & 154 (20.7\%) \\ 
   &  &  &  &  &  &  &  \\ 
  first & 10\% & 5\% & 35 (19.7\%) & 91 (23.2\%) & 79 (15.5\%) & 167 (20\%) & 53 (7.1\%) \\ 
  first & 10\% & 10\% & 41 (23\%) & 104 (26.5\%) & 82 (16.1\%) & 202 (24.2\%) & 76 (10.2\%) \\ 
  first & 10\% & 15\% & 50 (28.1\%) & 104 (26.5\%) & 85 (16.7\%) & 211 (25.3\%) & 85 (11.4\%) \\ 
  last & 10\% & 5\% & 65 (36.5\%) & 158 (40.2\%) & 177 (34.7\%) & 275 (33\%) & 105 (14.1\%) \\ 
  last & 10\% & 10\% & 77 (43.3\%) & 171 (43.5\%) & 181 (35.5\%) & 290 (34.8\%) & 145 (19.5\%) \\ 
  last & 10\% & 15\% & 79 (44.4\%) & 178 (45.3\%) & 184 (36.1\%) & 303 (36.4\%) & 157 (21.1\%) \\ 
   &  &  &  &  &  &  &  \\ 
   \bottomrule
\end{tabular}
\end{table}

\clearpage
\begin{landscape}
\begin{table}[p]
\scriptsize
\centering
\caption{
\textbf{T-Test comparing post-criticism productivity growth of treatment authors against matched population.}
``Authorship'' denotes the position of the authorship, either the first author or the last author. 
``Impact $\pm \epsilon$'' specifies the percentage tolerance for matching authors' citation impact, calculated as the log of their average 3-year field normalized impact of all papers published in the five years preceding receipt of the critical letter, normalized by the average across all authors in the same journal. 
``Productivity $\pm \epsilon$'' specifies the percentage tolerance for matching authors' productivity, calculated as their publication count, fractionalized by total authorship, normalized by the average across all authors in the same journal.
The T-Test compares the \textit{productivity growth} between the two populations. 
An authors' productivity growth is the ratio of their normalized and fractionalized productivity in the five years after receipt of the critical letter as compared to the five years before. 
In each cell, the first value indicates the test statistic of a paired T-Test.
The parenthetical next to the test statistic lists the p-value obtained from the test. 
}
\label{si:tab:authors-matched-productivity-ttest}
\begin{tabular}{llllllll}
  \toprule
Authorship & Impact $\pm$ $\epsilon$ & Productivity $\pm$ $\epsilon$ & Nature & Science & PNAS & PRL & Other APS \\ 
  \midrule
first & 5\% & 10\% & -0.22 (p=0.008) & -0.06 (p=0.496) & -0.06 (p=0.482) & -0.09 (p=0.177) & -0.04 (p=0.759) \\ 
  first & 10\% & 10\% & -0.06 (p=0.535) & -0.04 (p=0.590) & -0.03 (p=0.736) & -0.06 (p=0.402) & -0.09 (p=0.315) \\ 
  first & 15\% & 10\% & -0.09 (p=0.368) & -0.02 (p=0.756) & -0.12 (p=0.071) & -0.14 (p=0.044) & -0.02 (p=0.846) \\ 
  last & 5\% & 10\% & -0.03 (p=0.658) & -0.02 (p=0.718) & -0.07 (p=0.106) & -0.08 (p=0.017) & -0.1 (p=0.151) \\ 
  last & 10\% & 10\% & -0.09 (p=0.124) & -0.06 (p=0.217) & -0.06 (p=0.143) & -0.05 (p=0.134) & -0.13 (p=0.020) \\ 
  last & 15\% & 10\% & -0.24 (p=0.188) & -0.03 (p=0.531) & -0.11 (p=0.018) & -0.06 (p=0.052) & -0.11 (p=0.047) \\ 
   &  &  &  &  &  &  &  \\ 
  first & 10\% & 5\% & 0.08 (p=0.604) & -0.03 (p=0.668) & -0.05 (p=0.570) & -0.11 (p=0.139) & -0.13 (p=0.256) \\ 
  first & 10\% & 10\% & -0.06 (p=0.535) & -0.04 (p=0.590) & -0.03 (p=0.736) & -0.06 (p=0.402) & -0.09 (p=0.315) \\ 
  first & 10\% & 15\% & 0.01 (p=0.949) & 0.01 (p=0.839) & -0.05 (p=0.541) & -0.01 (p=0.877) & -0.07 (p=0.426) \\ 
  last & 10\% & 5\% & -0.11 (p=0.088) & -0.05 (p=0.329) & -0.1 (p=0.019) & -0.06 (p=0.108) & -0.08 (p=0.088) \\ 
  last & 10\% & 10\% & -0.09 (p=0.124) & -0.06 (p=0.217) & -0.06 (p=0.143) & -0.05 (p=0.134) & -0.13 (p=0.020) \\ 
  last & 10\% & 15\% & -0.06 (p=0.356) & -0.07 (p=0.147) & -0.04 (p=0.381) & -0.07 (p=0.043) & -0.12 (p=0.011) \\ 
   &  &  &  &  &  &  &  \\ 
   \bottomrule
\end{tabular}
\end{table}
\end{landscape}

\clearpage
\begin{landscape}
\begin{table}[p]
\scriptsize
\centering
\caption{
\textbf{T-Test comparing post-criticism impact growth of treatment authors against matched population.}
``Authorship'' denotes the position of the authorship, either the first author or the last author. 
``Impact $\pm \epsilon$'' specifies the percentage tolerance for matching authors' citation impact, calculated as the log of their average 3-year field normalized impact of all papers published in the five years preceding receipt of the critical letter, normalized by the average across all authors in the same journal. 
``Productivity $\pm \epsilon$'' specifies the percentage tolerance for matching authors' productivity, calculated as their publication count, fractionalized by total authorship, normalized by the average across all authors in the same journal.
The T-Test compares the \textit{impact growth} between the two populations. 
An authors' impact growth is the ratio of their average field-normalized 3-year citation impact of papers published five years after receipt of the critical letter compared to the five years before.
In each cell, the first value indicates the test statistic of a paired T-Test.
The parenthetical next to the test statistic lists the p-value obtained from the test. 
}
\label{si:tab:authors-matched-impact-ttest}
\centering
\begin{tabular}{llllllll}
  \toprule
Authorship & Impact $\pm$ $\epsilon$ & Productivity $\pm$ $\epsilon$ & Nature & Science & PNAS & PRL & Other APS \\ 
  \midrule
first & 5\% & 10\% & -0.13 (p=0.282) & 0.01 (p=0.921) & -0.07 (p=0.515) & 0.33 (p=0.238) & 0.06 (p=0.829) \\ 
  first & 10\% & 10\% & 0.02 (p=0.831) & 0.01 (p=0.823) & -0.12 (p=0.310) & 0.31 (p=0.216) & 0.09 (p=0.642) \\ 
  first & 15\% & 10\% & 0.06 (p=0.528) & 0.01 (p=0.854) & -0.08 (p=0.437) & 0.11 (p=0.179) & 0.04 (p=0.802) \\ 
  last & 5\% & 10\% & 0.1 (p=0.310) & 0.03 (p=0.768) & -0.09 (p=0.087) & -0.03 (p=0.544) & -0.02 (p=0.840) \\ 
  last & 10\% & 10\% & 0.03 (p=0.749) & -0.05 (p=0.445) & -0.07 (p=0.140) & -0.02 (p=0.568) & 0.03 (p=0.713) \\ 
  last & 15\% & 10\% & 0.04 (p=0.653) & -0.06 (p=0.347) & -0.04 (p=0.440) & -0.02 (p=0.665) & -0.05 (p=0.427) \\ 
   &  &  &  &  &  &  &  \\ 
  first & 10\% & 5\% & -0.06 (p=0.536) & 0.03 (p=0.671) & -0.11 (p=0.340) & 0.4 (p=0.180) & 0 (p=0.991) \\ 
  first & 10\% & 10\% & 0.02 (p=0.831) & 0.01 (p=0.823) & -0.12 (p=0.310) & 0.31 (p=0.216) & 0.09 (p=0.642) \\ 
  first & 10\% & 15\% & -0.03 (p=0.777) & 0.06 (p=0.480) & -0.19 (p=0.077) & 0.29 (p=0.239) & 0.07 (p=0.685) \\ 
  last & 10\% & 5\% & 0.03 (p=0.727) & -0.07 (p=0.367) & -0.05 (p=0.281) & -0.01 (p=0.818) & 0.04 (p=0.615) \\ 
  last & 10\% & 10\% & 0.03 (p=0.749) & -0.05 (p=0.445) & -0.07 (p=0.140) & -0.02 (p=0.568) & 0.03 (p=0.713) \\ 
  last & 10\% & 15\% & 0.09 (p=0.296) & -0.05 (p=0.397) & -0.04 (p=0.433) & 0.01 (p=0.824) & 0.05 (p=0.478) \\ 
   &  &  &  &  &  &  &  \\ 
   \bottomrule
\end{tabular}
\end{table}
\end{landscape}

\clearpage
\begin{landscape}
\begin{table}[p]
\centering
\caption{
  \textbf{Replies have lower visibility than their associated critical letters.}
  Shown are the average citation impact and number of tweets a subset of letters and replies in which links could be reliably identified. 
  Replies are identified from among the citations to letters based on whether the key word "reply" appears in its title. 
  This process could not identify replies in the journal \textit{Science}, and so this journal is excluded. 
  Moreover, counts are lower than the total number of letters as records that could not be matched to a reply are excluded.
  Common reasons are that there were multiple citing papers containing the term ``reply'' in their title, that the reply did not in fact use the term, or that the replies were not indexed in MAG. 
  Results illustrate that, across every journal, replies have lower visibility than the letters they are associated with.
}
\label{si:tab:letters-and-replies-visibility}
\begin{tabular}{lrrrrr}
  \toprule
Journal & N & Letter 3-year Impact & Reply 3-year Impact & Letter \#Tweets & Reply \#Tweets \\ 
  \midrule
Nature & 126 & 25.97 & 4.37 & 5.52 & 2.64 \\ 
  PNAS & 229 & 9.46 & 2.88 & 1.83 & 0.59 \\ 
  PRL & 188 & 7.10 & 3.52 & 0.01 & 0.02 \\ 
  Other APS & 158 & 6.72 & 3.35 & 0.04 & 0.03 \\ 
   \bottomrule
\end{tabular}
\end{table}
\end{landscape}

\clearpage
\printbibliography{}

@inproceedings{sinha2015overview,
  title={An overview of microsoft academic service (mas) and applications},
  author={Sinha, Arnab and Shen, Zhihong and Song, Yang and Ma, Hao and Eide, Darrin and Hsu, Bo-June and Wang, Kuansan},
  booktitle={Proceedings of the 24th international conference on world wide web},
  pages={243--246},
  year={2015}
}

@article{decruz_2013_value,
	title = {The value of epistemic disagreement in scientific practice. {The} case of {Homo} floresiensis},
	volume = {44},
	doi = {10.1016/j.shpsa.2013.02.002},
	number = {2},
	journal = {Studies in History and Philosophy of Science Part A},
	author = {Cruz, Helen De and Smedt, Johan De},
	month = jun,
	year = {2013},
	pages = {169--177},
}

@book{feyerabend_1975_against,
	address = {Atlantic Highlands, N.J.},
	title = {Against method: {Outline} of an anarchistic theory of knowledge},
	author = {Feyerabend, Paul},
	year = {1975},
}

@book{kitcher_advancement_1993,
	title = {The {Advancement} of {Science}: {Science} {Without} {Legend}, {Objectivity} {Without} {Illusions}},
	shorttitle = {The {Advancement} of {Science}},
	publisher = {Oxford University Press},
	author = {Kitcher, Philip},
	year = {1993},
}

@book{popper_logic_1959,
	title = {The {Logic} of {Scientific} {Discovery}},
	isbn = {978-1-61427-743-9},
	language = {English},
	publisher = {Martino Fine Books},
	author = {Popper, Karl R.},
	year = {1959},
}

@book{latour_science_1988,
	address = {Cambridge, Mass},
	title = {Science in {Action}: {How} to {Follow} {Scientists} and {Engineers} {Through} {Society}},
	isbn = {978-0-674-79291-3},
	shorttitle = {Science in {Action}},
	publisher = {Harvard University Press},
	author = {Latour, Bruno},
	month = oct,
	year = {1988},
}

@book{kumar_2010_quantum,
	address = {New York ; London},
	edition = {1st edition},
	title = {Quantum: {Einstein}, {Bohr}, and the {Great} {Debate} about the {Nature} of {Reality}},
	isbn = {978-0-393-07829-9},
	publisher = {W. W. Norton \& Company},
	author = {Kumar, Manjit},
	month = may,
	year = {2010},
}

@article{radicchi_2012_comment,
  title = {In science “there is no bad publicity”: Papers criticized in comments have high scientific impact},
  volume = {2},
  ISSN = {2045-2322},
  DOI = {10.1038/srep00815},
  number = {1},
  journal = {Scientific Reports},
  publisher = {Springer Science and Business Media LLC},
  author = {Radicchi,  Filippo},
  year = {2012},
  month = nov 
}

@article{catalini_2015_negative,
  title = {The incidence and role of negative citations in science},
  volume = {112},
  DOI = {10.1073/pnas.1502280112},
  number = {45},
  journal = {Proceedings of the National Academy of Sciences},
  publisher = {Proceedings of the National Academy of Sciences},
  author = {Catalini,  Christian and Lacetera,  Nicola and Oettl,  Alexander},
  year = {2015},
  month = oct,
  pages = {13823–13826}
}

@article{lamers_2021_disagreement,
  title = {Investigating disagreement in the scientific literature},
  volume = {10},
  DOI = {10.7554/elife.72737},
  journal = {eLife},
  publisher = {eLife Sciences Publications,  Ltd},
  author = {Lamers,  Wout S and Boyack,  Kevin and Larivière,  Vincent and Sugimoto,  Cassidy R and van Eck,  Nees Jan and Waltman,  Ludo and Murray,  Dakota},
  year = {2021},
  month = dec 
}

@article{cook_2013_quantifying,
  title = {Quantifying the consensus on anthropogenic global warming in the
               scientific literature},
  volume = {8},
  ISSN = {1748-9326},
  url = {http://dx.doi.org/10.1088/1748-9326/8/2/024024},
  DOI = {10.1088/1748-9326/8/2/024024},
  number = {2},
  journal = {Environmental Research Letters},
  publisher = {IOP Publishing},
  author = {Cook,  John and Nuccitelli,  Dana and Green,  Sarah A and Richardson,  Mark and Winkler,  B\"{a}rbel and Painting,  Rob and Way,  Robert and Jacobs,  Peter and Skuce,  Andrew},
  year = {2013},
  month = may,
  pages = {024024}
}

@article{azoulay_2017_scandal,
  title = {The career effects of scandal: Evidence from scientific retractions},
  volume = {46},
  DOI = {10.1016/j.respol.2017.07.003},
  number = {9},
  journal = {Research Policy},
  publisher = {Elsevier BV},
  author = {Azoulay,  Pierre and Bonatti,  Alessandro and Krieger,  Joshua L.},
  year = {2017},
  pages = {1552–1569}
}

@article{bastian2014stronger,
  title = {A Stronger Post-Publication Culture Is Needed for Better Science},
  volume = {11},
  ISSN = {1549-1676},
  url = {http://dx.doi.org/10.1371/journal.pmed.1001772},
  DOI = {10.1371/journal.pmed.1001772},
  number = {12},
  journal = {PLoS Medicine},
  publisher = {Public Library of Science (PLoS)},
  author = {Bastian,  Hilda},
  year = {2014},
  month = dec,
  pages = {e1001772}
}

@article{winker2015promise,
  title={The promise of post‐publication peer review: how do we get there from here?},
  author={Margaret A. Winker},
  journal={Learned Publishing},
  year={2015},
  volume={28},
  url={https://api.semanticscholar.org/CorpusID:206007497}
}

@article{hardwicke_2022_critique,
  title = {Post-publication critique at top-ranked journals across scientific disciplines: a cross-sectional assessment of policies and practice},
  volume = {9},
  ISSN = {2054-5703},
  url = {http://dx.doi.org/10.1098/rsos.220139},
  DOI = {10.1098/rsos.220139},
  number = {8},
  journal = {Royal Society Open Science},
  publisher = {The Royal Society},
  author = {Hardwicke,  Tom E. and Thibault,  Robert T. and Kosie,  Jessica E. and Tzavella,  Loukia and Bendixen,  Theiss and Handcock,  Sarah A. and K\"{o}neke,  Vivian E. and Ioannidis,  John P. A.},
  year = {2022},
  month = aug 
}

@article{peng2022retracted,
  title = {Dynamics of cross-platform attention to retracted papers},
  volume = {119},
  ISSN = {1091-6490},
  DOI = {10.1073/pnas.2119086119},
  number = {25},
  journal = {Proceedings of the National Academy of Sciences},
  publisher = {Proceedings of the National Academy of Sciences},
  author = {Peng,  Hao and Romero,  Daniel M. and Horvát,  Em\H{o}ke-\'{A}gnes},
  year = {2022}
}

@article{cokol_2007_retracted,
  title = {How many scientific papers should be retracted?},
  volume = {8},
  ISSN = {1469-3178},
  url = {http://dx.doi.org/10.1038/sj.embor.7400970},
  DOI = {10.1038/sj.embor.7400970},
  number = {5},
  journal = {EMBO reports},
  publisher = {Springer Science and Business Media LLC},
  author = {Cokol,  Murat and Iossifov,  Ivan and Rodriguez‐Esteban,  Raul and Rzhetsky,  Andrey},
  year = {2007},
  month = mar,
  pages = {422–423}
}

@book{Kuhn1977,
  title = {The Essential Tension: Selected Studies in Scientific Tradition and Change},
  ISBN = {9780226217239},
  url = {http://dx.doi.org/10.7208/chicago/9780226217239.001.0001},
  DOI = {10.7208/chicago/9780226217239.001.0001},
  publisher = {University of Chicago Press},
  author = {Kuhn,  Thomas S.},
  year = {1977}
}

@article{azoulay_2019_funeral,
  title = {Does Science Advance One Funeral at a Time?},
  volume = {109},
  ISSN = {0002-8282},
  url = {http://dx.doi.org/10.1257/aer.20161574},
  DOI = {10.1257/aer.20161574},
  number = {8},
  journal = {American Economic Review},
  publisher = {American Economic Association},
  author = {Azoulay,  Pierre and Fons-Rosen,  Christian and Zivin,  Joshua S. Graff},
  year = {2019},
  month = aug,
  pages = {2889–2920}
}

@article{wang_2017_novelty,
  title = {Bias against novelty in science: A cautionary tale for users of bibliometric indicators},
  volume = {46},
  ISSN = {0048-7333},
  url = {http://dx.doi.org/10.1016/j.respol.2017.06.006},
  DOI = {10.1016/j.respol.2017.06.006},
  number = {8},
  journal = {Research Policy},
  publisher = {Elsevier BV},
  author = {Wang,  Jian and Veugelers,  Reinhilde and Stephan,  Paula},
  year = {2017},
  month = oct,
  pages = {1416–1436}
}

@article{boudreau_2016_looking,
  title = {Looking Across and Looking Beyond the Knowledge Frontier: Intellectual Distance,  Novelty,  and Resource Allocation in Science},
  volume = {62},
  ISSN = {1526-5501},
  url = {http://dx.doi.org/10.1287/mnsc.2015.2285},
  DOI = {10.1287/mnsc.2015.2285},
  number = {10},
  journal = {Management Science},
  publisher = {Institute for Operations Research and the Management Sciences (INFORMS)},
  author = {Boudreau,  Kevin J. and Guinan,  Eva C. and Lakhani,  Karim R. and Riedl,  Christoph},
  year = {2016},
  month = sep,
  pages = {2765–2783}
}

@article{li_2017_expertise,
  title = {Expertise versus Bias in Evaluation: Evidence from the NIH},
  volume = {9},
  ISSN = {1945-7790},
  url = {http://dx.doi.org/10.1257/app.20150421},
  DOI = {10.1257/app.20150421},
  number = {2},
  journal = {American Economic Journal: Applied Economics},
  publisher = {American Economic Association},
  author = {Li,  Danielle},
  year = {2017},
  month = apr,
  pages = {60–92}
}

@article{siler_2016_review,
  title = {Peer Review and Scholarly Originality: Let 1, 000 Flowers Bloom,  but Don’t Step on Any},
  volume = {42},
  ISSN = {1552-8251},
  url = {http://dx.doi.org/10.1177/0162243916656919},
  DOI = {10.1177/0162243916656919},
  number = {1},
  journal = {Science,  Technology,  \& Human Values},
  publisher = {SAGE Publications},
  author = {Siler,  Kyle and Strang,  David},
  year = {2016},
  month = aug,
  pages = {29–61}
}

@article{lane_2022_conservatism,
  title = {Conservatism Gets Funded? A Field Experiment on the Role of Negative Information in Novel Project Evaluation},
  volume = {68},
  ISSN = {1526-5501},
  url = {http://dx.doi.org/10.1287/mnsc.2021.4107},
  DOI = {10.1287/mnsc.2021.4107},
  number = {6},
  journal = {Management Science},
  publisher = {Institute for Operations Research and the Management Sciences (INFORMS)},
  author = {Lane,  Jacqueline N. and Teplitskiy,  Misha and Gray,  Gary and Ranu,  Hardeep and Menietti,  Michael and Guinan,  Eva C. and Lakhani,  Karim R.},
  year = {2022},
  month = jun,
  pages = {4478–4495}
}

@article{ayoubi_2021_novel,
  title = {Does It Pay to Do Novel Science? The Selectivity Patterns in Science Funding},
  volume = {48},
  ISSN = {1471-5430},
  url = {http://dx.doi.org/10.1093/scipol/scab031},
  DOI = {10.1093/scipol/scab031},
  number = {5},
  journal = {Science and Public Policy},
  publisher = {Oxford University Press (OUP)},
  author = {Ayoubi,  Charles and Pezzoni,  Michele and Visentin,  Fabiana},
  year = {2021},
  month = jun,
  pages = {635–648}
}

@article{teplitskiy_2022_novel,
  title = {Is novel research worth doing? Evidence from peer review at 49 journals},
  volume = {119},
  ISSN = {1091-6490},
  url = {http://dx.doi.org/10.1073/pnas.2118046119},
  DOI = {10.1073/pnas.2118046119},
  number = {47},
  journal = {Proceedings of the National Academy of Sciences},
  publisher = {Proceedings of the National Academy of Sciences},
  author = {Teplitskiy,  Misha and Peng,  Hao and Blasco,  Andrea and Lakhani,  Karim R.},
  year = {2022},
  month = nov 
}

@article{uzzi_2013_atypical,
  title = {Atypical Combinations and Scientific Impact},
  volume = {342},
  ISSN = {1095-9203},
  url = {http://dx.doi.org/10.1126/science.1240474},
  DOI = {10.1126/science.1240474},
  number = {6157},
  journal = {Science},
  publisher = {American Association for the Advancement of Science (AAAS)},
  author = {Uzzi,  Brian and Mukherjee,  Satyam and Stringer,  Michael and Jones,  Ben},
  year = {2013},
  month = oct,
  pages = {468–472}
}

@article{brohman_2016_interdisciplinary,
  title = {Interdisciplinary research has consistently lower funding success},
  volume = {534},
  ISSN = {1476-4687},
  url = {http://dx.doi.org/10.1038/nature18315},
  DOI = {10.1038/nature18315},
  number = {7609},
  journal = {Nature},
  publisher = {Springer Science and Business Media LLC},
  author = {Bromham,  Lindell and Dinnage,  Russell and Hua,  Xia},
  year = {2016},
  month = jun,
  pages = {684–687}
}

@book{Klein2021,
  title = {Beyond Interdisciplinarity: Boundary Work,  Communication,  and Collaboration},
  ISBN = {9780197571187},
  url = {http://dx.doi.org/10.1093/oso/9780197571149.001.0001},
  DOI = {10.1093/oso/9780197571149.001.0001},
  publisher = {Oxford University PressNew York},
  author = {Klein,  Julie Thompson},
  year = {2021},
  month = aug 
}

@book{Gieryn1999,
  title     = "Cultural boundaries of science",
  author    = "Gieryn, Thomas F",
  publisher = "University of Chicago Press",
  edition   =  2,
  month     =  feb,
  year      =  1999,
  address   = "Chicago, IL"
}

@article{yegros-yegros_2015_interdisciplinarity,
  title = {Does Interdisciplinary Research Lead to Higher Citation Impact? The Different Effect of Proximal and Distal Interdisciplinarity},
  volume = {10},
  ISSN = {1932-6203},
  url = {http://dx.doi.org/10.1371/journal.pone.0135095},
  DOI = {10.1371/journal.pone.0135095},
  number = {8},
  journal = {PLOS ONE},
  publisher = {Public Library of Science (PLoS)},
  author = {Yegros-Yegros,  Alfredo and Rafols,  Ismael and D’Este,  Pablo},
  editor = {Glanzel,  Wolfgang},
  year = {2015},
  month = aug,
  pages = {e0135095}
}

@article{chen_2015_interdisciplinary,
  title = {Are top-cited papers more interdisciplinary?},
  volume = {9},
  ISSN = {1751-1577},
  url = {http://dx.doi.org/10.1016/j.joi.2015.09.003},
  DOI = {10.1016/j.joi.2015.09.003},
  number = {4},
  journal = {Journal of Informetrics},
  publisher = {Elsevier BV},
  author = {Chen,  Shiji and Arsenault,  Clément and Larivière,  Vincent},
  year = {2015},
  month = oct,
  pages = {1034–1046}
}

@article{azoulay_2015_retractions,
  title = {Retractions},
  volume = {97},
  ISSN = {1530-9142},
  url = {http://dx.doi.org/10.1162/REST_a_00469},
  DOI = {10.1162/rest_a_00469},
  number = {5},
  journal = {Review of Economics and Statistics},
  publisher = {MIT Press - Journals},
  author = {Azoulay,  Pierre and Furman,  Jeffrey L. and Krieger and Murray,  Fiona},
  year = {2015},
  month = dec,
  pages = {1118–1136}
}

@article{berger_2010_publicity,
  title = {Positive Effects of Negative Publicity: When Negative Reviews Increase Sales},
  volume = {29},
  ISSN = {1526-548X},
  url = {http://dx.doi.org/10.1287/mksc.1090.0557},
  DOI = {10.1287/mksc.1090.0557},
  number = {5},
  journal = {Marketing Science},
  publisher = {Institute for Operations Research and the Management Sciences (INFORMS)},
  author = {Berger,  Jonah and Sorensen,  Alan T. and Rasmussen,  Scott J.},
  year = {2010},
  month = sep,
  pages = {815–827}
}

@book{oreskes_2010_merchants,
  title     = "Merchants of doubt",
  author    = "Oreskes, Naomi and Conway, Erik M",
  publisher = "Bloomsbury Press",
  month     =  may,
  year      =  2010,
  address   = "New York, NY"
}

@article{lewandowsky_2012_consensus,
  title = {The pivotal role of perceived scientific consensus in acceptance of science},
  volume = {3},
  ISSN = {1758-6798},
  url = {http://dx.doi.org/10.1038/nclimate1720},
  DOI = {10.1038/nclimate1720},
  number = {4},
  journal = {Nature Climate Change},
  publisher = {Springer Science and Business Media LLC},
  author = {Lewandowsky,  Stephan and Gignac,  Gilles E. and Vaughan,  Samuel},
  year = {2012},
  month = oct,
  pages = {399–404}
}

@article{christensen_2007_disagreement,
  title = {Epistemology of Disagreement: The Good News},
  volume = {116},
  ISSN = {1558-1470},
  url = {http://dx.doi.org/10.1215/00318108-2006-035},
  DOI = {10.1215/00318108-2006-035},
  number = {2},
  journal = {The Philosophical Review},
  publisher = {Duke University Press},
  author = {Christensen,  David},
  year = {2007},
  month = apr,
  pages = {187–217}
}

@article{schneider_2020_retraction,
  title = {Continued post-retraction citation of a fraudulent clinical trial report,  11 years after it was retracted for falsifying data},
  volume = {125},
  ISSN = {1588-2861},
  url = {http://dx.doi.org/10.1007/s11192-020-03631-1},
  DOI = {10.1007/s11192-020-03631-1},
  number = {3},
  journal = {Scientometrics},
  publisher = {Springer Science and Business Media LLC},
  author = {Schneider,  Jodi and Ye,  Di and Hill,  Alison M. and Whitehorn,  Ashley S.},
  year = {2020},
  month = oct,
  pages = {2877–2913}
}

@article{mcdiarmid_2021_replication,
  title = {Psychologists update their beliefs about effect sizes after replication studies},
  volume = {5},
  ISSN = {2397-3374},
  url = {http://dx.doi.org/10.1038/s41562-021-01220-7},
  DOI = {10.1038/s41562-021-01220-7},
  number = {12},
  journal = {Nature Human Behaviour},
  publisher = {Springer Science and Business Media LLC},
  author = {McDiarmid,  Alex D. and Tullett,  Alexa M. and Whitt,  Cassie M. and Vazire,  Simine and Smaldino,  Paul E. and Stephens,  Jeremy E.},
  year = {2021},
  month = nov,
  pages = {1663–1673}
}

@article{ortega2021pubpeer,
  title = {Classification and analysis of PubPeer comments: How a web journal club is used},
  volume = {73},
  ISSN = {2330-1643},
  url = {http://dx.doi.org/10.1002/asi.24568},
  DOI = {10.1002/asi.24568},
  number = {5},
  journal = {Journal of the Association for Information Science and Technology},
  publisher = {Wiley},
  author = {Ortega,  José Luis},
  year = {2021},
  month = aug,
  pages = {655–670}
}

@article{mcleish_2016_interdisciplinary,
  title = {Evaluating interdisciplinary research: the elephant in the peer-reviewers’ room},
  volume = {2},
  ISSN = {2055-1045},
  url = {http://dx.doi.org/10.1057/palcomms.2016.55},
  DOI = {10.1057/palcomms.2016.55},
  number = {1},
  journal = {Palgrave Communications},
  publisher = {Springer Science and Business Media LLC},
  author = {McLeish,  Tom and Strang,  Veronica},
  year = {2016},
  month = aug 
}

@article{funk_2017_disruption,
  title = {A Dynamic Network Measure of Technological Change},
  volume = {63},
  ISSN = {1526-5501},
  url = {http://dx.doi.org/10.1287/mnsc.2015.2366},
  DOI = {10.1287/mnsc.2015.2366},
  number = {3},
  journal = {Management Science},
  publisher = {Institute for Operations Research and the Management Sciences (INFORMS)},
  author = {Funk,  Russell J. and Owen-Smith,  Jason},
  year = {2017},
  month = mar,
  pages = {791–817}
}

@misc{leibel_2024_disruption,
    Author = {Christian Leibel and Lutz Bornmann},
    Title = {The disruption index in the multiverse: The calculation of scores comes with numerous (hidden) degrees of freedom},
    Year = {2024},
    Eprint = {arXiv:2406.13367},
}

@article{serghiou_2021_attention,
  title = {Media and social media attention to retracted articles according to Altmetric},
  volume = {16},
  ISSN = {1932-6203},
  url = {http://dx.doi.org/10.1371/journal.pone.0248625},
  DOI = {10.1371/journal.pone.0248625},
  number = {5},
  journal = {PLOS ONE},
  publisher = {Public Library of Science (PLoS)},
  author = {Serghiou,  Stylianos and Marton,  Rebecca M. and Ioannidis,  John P. A.},
  editor = {Dorta-González,  Pablo},
  year = {2021},
  month = may,
  pages = {e0248625}
}

@article{lariviere_2016_division,
    author = {Vincent Larivière and Nadine Desrochers and Benoît Macaluso and Philippe Mongeon and Adèle Paul-Hus and Cassidy R Sugimoto},
    title ={Contributorship and division of labor in knowledge production},
    journal = {Social Studies of Science},
    volume = {46},
    number = {3},
    pages = {417-435},
    year = {2016},
    doi = {10.1177/0306312716650046}
}

@article{lariviere_2021_division,
    author = {Larivière, Vincent and Pontille, David and Sugimoto, Cassidy R.},
    title = "{Investigating the division of scientific labor using the Contributor
                    Roles Taxonomy (CRediT)}",
    journal = {Quantitative Science Studies},
    volume = {2},
    number = {1},
    pages = {111-128},
    year = {2021},
    month = {04},
    doi = {10.1162/qss_a_00097}
}

@article{matchit_2011,
    title = {{MatchIt}: Nonparametric Preprocessing for Parametric
      Causal Inference},
    author = {Daniel E. Ho and Kosuke Imai and Gary King and Elizabeth
      A. Stuart},
    year = {2011},
    journal = {Journal of Statistical Software},
    volume = {42},
    number = {8},
    pages = {1--28},
    doi = {10.18637/jss.v042.i08},
  }

@article{simpson_1949_diversity,
  title = {Measurement of Diversity},
  volume = {163},
  ISSN = {1476-4687},
  url = {http://dx.doi.org/10.1038/163688a0},
  DOI = {10.1038/163688a0},
  number = {4148},
  journal = {Nature},
  publisher = {Springer Science and Business Media LLC},
  author = {SIMPSON,  E. H.},
  year = {1949},
  month = apr,
  pages = {688–688}
}

@article{lin_2023_sciscinet,
  title = {SciSciNet: A large-scale open data lake for the science of science research},
  volume = {10},
  ISSN = {2052-4463},
  url = {http://dx.doi.org/10.1038/s41597-023-02198-9},
  DOI = {10.1038/s41597-023-02198-9},
  number = {1},
  journal = {Scientific Data},
  publisher = {Springer Science and Business Media LLC},
  author = {Lin,  Zihang and Yin,  Yian and Liu,  Lu and Wang,  Dashun},
  year = {2023},
  month = jun 
}

@article{huang_2020_gender,
  title = {Historical comparison of gender inequality in scientific careers across countries and disciplines},
  volume = {117},
  ISSN = {1091-6490},
  DOI = {10.1073/pnas.1914221117},
  number = {9},
  journal = {Proceedings of the National Academy of Sciences},
  publisher = {Proceedings of the National Academy of Sciences},
  author = {Huang,  Junming and Gates,  Alexander J. and Sinatra,  Roberta and Barabási,  Albert-László},
  year = {2020},
  month = feb,
  pages = {4609–4616}
}

@dataset{van_eck_2023_leiden,
  author       = {Van Eck, Nees Jan},
  title        = {CWTS Leiden Ranking 2023},
  month        = jun,
  year         = 2023,
  publisher    = {Zenodo},
  doi          = {10.5281/zenodo.8027120},
  url          = {https://doi.org/10.5281/zenodo.8027120}
}

@article{hengel_2022_female,
  title = {Publishing While Female: are Women Held to Higher Standards? Evidence from Peer Review},
  volume = {132},
  ISSN = {1468-0297},
  url = {http://dx.doi.org/10.1093/ej/ueac032},
  DOI = {10.1093/ej/ueac032},
  number = {648},
  journal = {The Economic Journal},
  publisher = {Oxford University Press (OUP)},
  author = {Hengel,  Erin},
  year = {2022},
  month = may,
  pages = {2951–2991}
}

@article{fox_2019_gender,
  title = {Gender differences in peer review outcomes and manuscript impact at six journals of ecology and evolution},
  volume = {9},
  ISSN = {2045-7758},
  url = {http://dx.doi.org/10.1002/ece3.4993},
  DOI = {10.1002/ece3.4993},
  number = {6},
  journal = {Ecology and Evolution},
  publisher = {Wiley},
  author = {Fox,  Charles W. and Paine,  C. E. Timothy},
  year = {2019},
  month = mar,
  pages = {3599–3619}
}

@article{moss-racusin_2012_bias,
  title = {Science faculty’s subtle gender biases favor male students},
  volume = {109},
  ISSN = {1091-6490},
  url = {http://dx.doi.org/10.1073/pnas.1211286109},
  DOI = {10.1073/pnas.1211286109},
  number = {41},
  journal = {Proceedings of the National Academy of Sciences},
  publisher = {Proceedings of the National Academy of Sciences},
  author = {Moss-Racusin,  Corinne A. and Dovidio,  John F. and Brescoll,  Victoria L. and Graham,  Mark J. and Handelsman,  Jo},
  year = {2012},
  month = sep,
  pages = {16474–16479}
}

@inproceedings{Singh2022SciRepEvalAM,
  title={SciRepEval: A Multi-Format Benchmark for Scientific Document Representations},
  author={Amanpreet Singh and Mike D'Arcy and Arman Cohan and Doug Downey and Sergey Feldman},
  booktitle={Conference on Empirical Methods in Natural Language Processing},
  year={2022}
}

@inproceedings{cohan-etal-2020-specter,
    title = "{SPECTER}: Document-level Representation Learning using Citation-informed Transformers",
    author = "Cohan, Arman  and
      Feldman, Sergey  and
      Beltagy, Iz  and
      Downey, Doug  and
      Weld, Daniel",
    booktitle = "Proceedings of the 58th Annual Meeting of the Association for Computational Linguistics",
    month = jul,
    year = "2020",
    doi = "10.18653/v1/2020.acl-main.207",
    pages = "2270--2282"
}

@inproceedings{beltagy-etal-2019-scibert,
    title = "{S}ci{BERT}: A Pretrained Language Model for Scientific Text",
    author = "Beltagy, Iz  and
      Lo, Kyle  and
      Cohan, Arman",
    booktitle = "Proceedings of the 2019 Conference on Empirical Methods in Natural Language Processing and the 9th International Joint Conference on Natural Language Processing (EMNLP-IJCNLP)",
    month = nov,
    year = "2019",
    address = "Hong Kong, China",
    doi = "10.18653/v1/D19-1371",
    pages = "3615--3620",
}

@article{stang_2018_quotation,
  title = {Case study in major quotation errors: a critical commentary on the Newcastle–Ottawa scale},
  volume = {33},
  ISSN = {1573-7284},
  url = {http://dx.doi.org/10.1007/s10654-018-0443-3},
  DOI = {10.1007/s10654-018-0443-3},
  number = {11},
  journal = {European Journal of Epidemiology},
  publisher = {Springer Science and Business Media LLC},
  author = {Stang,  Andreas and Jonas,  Stephan and Poole,  Charles},
  year = {2018},
  month = sep,
  pages = {1025–1031}
}

@article{woo_2024_retractions,
  title = {On the shoulders of fallen giants: What do references to retracted research tell us about citation behaviors?},
  volume = {5},
  ISSN = {2641-3337},
  url = {http://dx.doi.org/10.1162/qss_a_00303},
  DOI = {10.1162/qss_a_00303},
  number = {1},
  journal = {Quantitative Science Studies},
  publisher = {MIT Press},
  author = {Woo,  Seokkyun and Walsh,  John P.},
  year = {2024},
  pages = {1–30}
}

@article{leung_2017_letter,
  title = {A 1980 Letter on the Risk of Opioid Addiction},
  volume = {376},
  ISSN = {1533-4406},
  url = {http://dx.doi.org/10.1056/NEJMc1700150},
  DOI = {10.1056/nejmc1700150},
  number = {22},
  journal = {New England Journal of Medicine},
  publisher = {Massachusetts Medical Society},
  author = {Leung,  Pamela T.M. and Macdonald,  Erin M. and Stanbrook,  Matthew B. and Dhalla,  Irfan A. and Juurlink,  David N.},
  year = {2017},
  month = jun,
  pages = {2194–2195}
}

@article{hsiao_2021_retracted,
  title = {Continued use of retracted papers: Temporal trends in citations and (lack of) awareness of retractions shown in citation contexts in biomedicine},
  volume = {2},
  ISSN = {2641-3337},
  url = {http://dx.doi.org/10.1162/qss_a_00155},
  DOI = {10.1162/qss_a_00155},
  number = {4},
  journal = {Quantitative Science Studies},
  publisher = {MIT Press},
  author = {Hsiao,  Tzu-Kun and Schneider,  Jodi},
  year = {2021},
  pages = {1144–1169}
}

@article{BarIlan_2017_post-retraction,
  title = {Post retraction citations in context: a case study},
  volume = {113},
  ISSN = {1588-2861},
  url = {http://dx.doi.org/10.1007/s11192-017-2242-0},
  DOI = {10.1007/s11192-017-2242-0},
  number = {1},
  journal = {Scientometrics},
  publisher = {Springer Science and Business Media LLC},
  author = {Bar-Ilan,  Judit and Halevi,  Gali},
  year = {2017},
  month = mar,
  pages = {547–565}
}

@article{BornemannCimenti_2015_retracted,
  title = {Perpetuation of Retracted Publications Using the Example of the Scott S. Reuben Case: Incidences,  Reasons and Possible Improvements},
  volume = {22},
  ISSN = {1471-5546},
  url = {http://dx.doi.org/10.1007/s11948-015-9680-y},
  DOI = {10.1007/s11948-015-9680-y},
  number = {4},
  journal = {Science and Engineering Ethics},
  publisher = {Springer Science and Business Media LLC},
  author = {Bornemann-Cimenti,  Helmar and Szilagyi,  Istvan S. and Sandner-Kiesling,  Andreas},
  year = {2015},
  month = jul,
  pages = {1063–1072}
}

@article{lakens_2020_critics,
  title = {Pandemic researchers — recruit your own best critics},
  volume = {581},
  ISSN = {1476-4687},
  DOI = {10.1038/d41586-020-01392-8},
  number = {7807},
  journal = {Nature},
  publisher = {Springer Science and Business Media LLC},
  author = {Lakens,  Daniël},
  year = {2020},
  month = may,
  pages = {121–121}
}

@article{deroover_2022_experts,
  title = {Why do experts disagree? The development of a taxonomy},
  volume = {32},
  ISSN = {1361-6609},
  url = {http://dx.doi.org/10.1177/09636625221110029},
  DOI = {10.1177/09636625221110029},
  number = {2},
  journal = {Public Understanding of Science},
  publisher = {SAGE Publications},
  author = {Deroover,  Kristine and Knight,  Simon and Burke,  Paul F. and Bucher,  Tamara},
  year = {2022},
  month = aug,
  pages = {224–246}
}

@article{schafmeister_2021_replication,
  title = {The Effect of Replications on Citation Patterns: Evidence From a Large-Scale Reproducibility Project},
  volume = {32},
  ISSN = {1467-9280},
  url = {http://dx.doi.org/10.1177/09567976211005767},
  DOI = {10.1177/09567976211005767},
  number = {10},
  journal = {Psychological Science},
  publisher = {SAGE Publications},
  author = {Schafmeister,  Felix},
  year = {2021},
  month = sep,
  pages = {1537–1548}
}

@article{hardwicke_2021_contradictory,
  title = {Citation Patterns Following a Strongly Contradictory Replication Result: Four           Case Studies From Psychology},
  volume = {4},
  ISSN = {2515-2467},
  url = {http://dx.doi.org/10.1177/25152459211040837},
  DOI = {10.1177/25152459211040837},
  number = {3},
  journal = {Advances in Methods and Practices in Psychological Science},
  publisher = {SAGE Publications},
  author = {Hardwicke,  Tom E. and Szűcs,  Dénes and Thibault,  Robert T. and Cr\"{u}well,  Sophia and van den Akker,  Olmo R. and Nuijten,  Michèle B. and Ioannidis,  John P. A.},
  year = {2021},
  month = jul 
}

@article{thelwall_2020_open,
  title = {Does the use of open,  non-anonymous peer review in scholarly publishing introduce bias? Evidence from the F1000Research post-publication open peer review publishing model},
  volume = {47},
  ISSN = {1741-6485},
  url = {http://dx.doi.org/10.1177/0165551520938678},
  DOI = {10.1177/0165551520938678},
  number = {6},
  journal = {Journal of Information Science},
  publisher = {SAGE Publications},
  author = {Thelwall,  Mike and Allen,  Liz and Papas,  Eleanor-Rose and Nyakoojo,  Zena and Weigert,  Verena},
  year = {2020},
  month = jul,
  pages = {809–820}
}

@article{soltesz_2020_interventions,
  title = {The effect of interventions on COVID-19},
  volume = {588},
  ISSN = {1476-4687},
  DOI = {10.1038/s41586-020-3025-y},
  number = {7839},
  journal = {Nature},
  publisher = {Springer Science and Business Media LLC},
  author = {Soltesz,  Kristian and Gustafsson,  Fredrik and Timpka,  Toomas and Jaldén,  Joakim and Jidling,  Carl and Heimerson,  Albin and Sch\"{o}n,  Thomas B. and Spreco,  Armin and Ekberg,  Joakim and Dahlstr\"{o}m,  \"{O}rjan and Bagge Carlson,  Fredrik and J\"{o}ud,  Anna and Bernhardsson,  Bo},
  year = {2020},
  month = dec,
  pages = {E26–E28}
}

@article{rosen_2019_temperature,
  title = {Temperature impact on GDP growth is overestimated},
  volume = {116},
  ISSN = {1091-6490},
  url = {http://dx.doi.org/10.1073/pnas.1908081116},
  DOI = {10.1073/pnas.1908081116},
  number = {33},
  journal = {Proceedings of the National Academy of Sciences},
  publisher = {Proceedings of the National Academy of Sciences},
  author = {Rosen,  Richard A.},
  year = {2019},
  month = aug,
  pages = {16170–16170}
}

@article{loisel_2019_soils,
  title = {Soils can help mitigate CO
            2
            emissions,  despite the challenges},
  volume = {116},
  ISSN = {1091-6490},
  url = {http://dx.doi.org/10.1073/pnas.1900444116},
  DOI = {10.1073/pnas.1900444116},
  number = {21},
  journal = {Proceedings of the National Academy of Sciences},
  publisher = {Proceedings of the National Academy of Sciences},
  author = {Loisel,  Julie and Casellas Connors,  John P. and Hugelius,  Gustaf and Harden,  Jennifer W. and Morgan,  Christine L.},
  year = {2019},
  month = may,
  pages = {10211–10212}
}

@article{rice_2020_rna,
  title = {Comment on “RNA-guided DNA insertion with CRISPR-associated transposases”},
  volume = {368},
  ISSN = {1095-9203},
  url = {http://dx.doi.org/10.1126/science.abb2022},
  DOI = {10.1126/science.abb2022},
  number = {6495},
  journal = {Science},
  publisher = {American Association for the Advancement of Science (AAAS)},
  author = {Rice,  Phoebe A. and Craig,  Nancy L. and Dyda,  Fred},
  year = {2020},
  month = jun 
}

@article{vandenHeuvel_2020_black-hole,
  title = {Comment on “A noninteracting low-mass black hole–giant star binary system”},
  volume = {368},
  ISSN = {1095-9203},
  url = {http://dx.doi.org/10.1126/science.aba3282},
  DOI = {10.1126/science.aba3282},
  number = {6491},
  journal = {Science},
  publisher = {American Association for the Advancement of Science (AAAS)},
  author = {van den Heuvel,  Ed P. J. and Tauris,  Thomas M.},
  year = {2020},
  month = may 
}

@article{HaibeKains_2020_transparency,
  title = {Transparency and reproducibility in artificial intelligence},
  volume = {586},
  ISSN = {1476-4687},
  url = {http://dx.doi.org/10.1038/s41586-020-2766-y},
  DOI = {10.1038/s41586-020-2766-y},
  number = {7829},
  journal = {Nature},
  publisher = {Springer Science and Business Media LLC},
  author = {Haibe-Kains,  Benjamin and Adam,  George Alexandru and Hosny,  Ahmed and Khodakarami,  Farnoosh and Shraddha,  Thakkar and Kusko,  Rebecca and Sansone,  Susanna-Assunta and Tong,  Weida and Wolfinger,  Russ D. and Mason,  Christopher E. and Jones,  Wendell and Dopazo,  Joaquin and Furlanello,  Cesare and Waldron,  Levi and Wang,  Bo and McIntosh,  Chris and Goldenberg,  Anna and Kundaje,  Anshul and Greene,  Casey S. and Broderick,  Tamara and Hoffman,  Michael M. and Leek,  Jeffrey T. and Korthauer,  Keegan and Huber,  Wolfgang and Brazma,  Alvis and Pineau,  Joelle and Tibshirani,  Robert and Hastie,  Trevor and Ioannidis,  John P. A. and Quackenbush,  John and Aerts,  Hugo J. W. L.},
  year = {2020},
  month = oct,
  pages = {E14–E16}
}
\end{refsection}

\end{document}